\documentclass[fleqn,usenatbib]{mnras}

\usepackage[T1]{fontenc}
\usepackage{ae,aecompl}

\usepackage{graphicx}	
\usepackage{amsmath}	
\usepackage{amssymb}	
\usepackage{float}      
\usepackage{diagbox}
\usepackage{gensymb}  

\usepackage{hyperref}
\usepackage{xspace}

\newcommand{\angstrom}{\mbox{\normalfont\AA}\xspace}

\newcommand{\se}{Squared Exponential\xspace}
\newcommand{\mtt}{Mat\'ern-3/2\xspace}
\newcommand{\mft}{Mat\'ern-5/2\xspace}

\newcommand{\mb}{\ensuremath{m_B^{\rm max}}\xspace}
\newcommand{\tmax}{$t^{\rm max}$\xspace}
\newcommand{\tmaxB}{$t^{\rm max}$\xspace}
\newcommand{\Mb}{M$_B^{\rm max}$\xspace}
\newcommand{\dm}{$\Delta$m$_{15}$($B$)\xspace}
\newcommand{\colour}{$(B - V)_{\rm max}$\xspace}

\newcommand{\pisco}{PISCOLA\xspace}

\newcommand{\orcid}[1]{\href{https://orcid.org/#1}{\includegraphics[width=10pt]{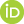}}}
\newcommand{\github}[1]{\href{https://github.com/#1}{\includegraphics[width=10pt]{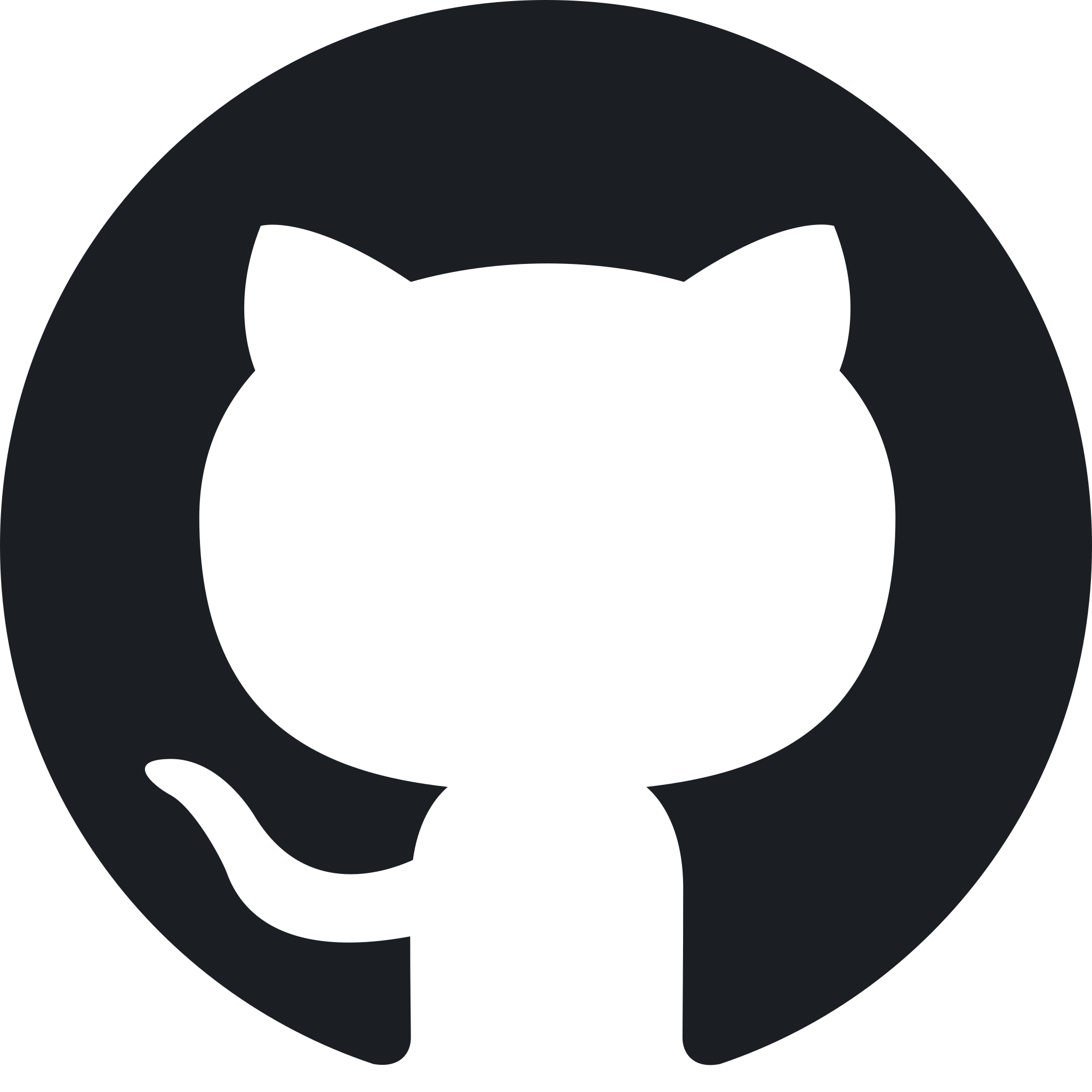}}}

\usepackage{newtxtext,newtxmath}

\title[\pisco]{PISCOLA: a data-driven transient light-curve fitter}

\author[M\"uller-Bravo et al.]{
Tom\'as E. M\"uller-Bravo$^{1}$\orcid{0000-0003-3939-7167}\thanks{E-mail: t.e.muller-bravo@soton.ac.uk},
Mark Sullivan$^{1}$\orcid{0000-0001-9053-4820},
Mathew Smith$^{1,2}$,
\newauthor
Chris Frohmaier$^{1}$\orcid{0000-0001-9553-4723},
Claudia P. Guti\'errez$^{1,3,4}$\orcid{0000-0003-2375-2064},
Philip Wiseman$^{1}$\orcid{0000-0002-3073-1512},
Zoe Zontou$^{1}$\\
$^{1}$School of Physics and Astronomy, University of Southampton, Southampton, Hampshire, SO17 1BJ, UK\\
$^{2}$Univ Lyon, Univ Claude Bernard Lyon 1, CNRS, IP2I Lyon / IN2P3, IMR 5822, F-69622, Villeurbanne, France\\
$^{3}$Finnish Centre for Astronomy with ESO (FINCA), FI-20014 University of Turku, Finland\\
$^{4}$Tuorla Observatory, Department of Physics and Astronomy, FI-20014 University of Turku, Finland\\
}

\date{Accepted XXX. Received YYY; in original form ZZZ}

\pubyear{2021}

\begin{document}
\label{firstpage}
\pagerange{\pageref{firstpage}--\pageref{lastpage}}
\maketitle

\begin{abstract}
Forthcoming time-domain surveys, such as the Rubin Observatory Legacy Survey of Space and Time, will vastly increase samples of supernovae (SNe) and other optical transients, requiring new data-driven techniques to analyse their photometric light curves. Here, we present the \lq Python for Intelligent Supernova-COsmology Light-curve Analysis\rq\ (\pisco), an open source data-driven light-curve fitter using Gaussian Processes that can estimate rest-frame light curves of transients without the need for an underlying light-curve template. We test \pisco\ on large-scale simulations of type Ia SNe (SNe Ia) to validate its performance, and show it successfully retrieves rest-frame peak magnitudes for average survey cadences of up to 7 days. We also compare to the existing SN Ia light-curve fitter SALT2 on real data, and find only small (but significant) disagreements for different light-curve parameters. As a proof-of-concept of an application of \pisco, we decomposed and analysed the \pisco rest-frame light-curves of SNe Ia from the Pantheon SN Ia sample with Non-Negative Matrix Factorization. Our new parametrization provides a similar performance to existing light-curve fitters such as SALT2. We further derived a SN Ia colour law from \pisco fits over $\sim$3500 to 7000\AA, and find agreement with the SALT2 colour law and with reddening laws with total-to-selective extinction ratio $R_V \lesssim 3.1$.
\github{temuller/piscola}
\end{abstract}

\begin{keywords}
software: data analysis -- supernovae: general -- cosmology: observations -- distance scale
\end{keywords}


\section{Introduction}
\label{sec:intro}

The next decade will see a rapid increase in the number and quality of extragalactic optical transients observed in wide-field, time-domain sky surveys. Detected objects will include classical supernova types, as well as new and exotic transients evolving on a wide-range of timescales. The scientific analyses of these transients will almost always require the fitting of the observed transient light-curves, followed by an estimation of their rest-frame properties, luminosities and timescales. Further, the study of the demographics of the populations these events are drawn from will require the automated fitting of many thousands of transients.

In this paper, we present \pisco, a data-driven transient light-curve fitter. Other than practical considerations such as ease of use, there are three main advantages of using \pisco for light-curve fitting. The first is that \pisco uses Gaussian processes for the light-curve fitting, and thus is not limited by an underlying light-curve model or template, and returns the intrinsic light-curve of a transient rather than a template fit. Secondly, other than an spectral energy distribution (SED) that is used for K-corrections, \pisco is agnostic to the type of transient being fit. Thus, it can in principle be used to estimate the rest-frame light-curves of any optical transient. Finally, as there is no underlying template, the intrinsic rest-frame light-curves/colours of the transient being fit can be measured across all phases of the transient's evolution: \pisco has no assumed colour evolution or colour law.

As a proof-of-concept of \pisco, in this paper we demonstrate the application of \pisco to type Ia supernovae (SNe Ia). SNe Ia are the result of a runaway thermonuclear explosion of carbon-oxygen white dwarf star \citep{Hoyle60, Woosley86}. The resulting light curves are powered by the radioactive decay of $^{56}$Ni \citep[e.g.,][]{Colgate69}, reaching a peak luminosity about 20 days after explosion \citep[e.g.,][]{1999AJ....118.2675R}. Although the exact triggering mechanism and configuration of their progenitor systems remains controversial \citep[see review of][]{2019NatAs...3..706J}, empirically SNe Ia appear similar in brightness with an intrinsic root-mean-square (r.m.s.) scatter in their peak $B$-band magnitudes of around 0.3\,mag \citep[e.g.,][]{Cadonau85, Tammann90}. Correlations between their peak luminosity, the decline rate of their light curves and their optical colours allow them to be further standardised \citep[e.g.,][]{Pskovskii77, Phillips93, Tripp98, Phillips99, Kattner12}, reducing their observed r.m.s. scatter to around $\sim$0.15\,mag or 7--8 per cent in distance \citep[e.g.,][]{Betoule14}. SNe Ia have therefore assumed a key role as cosmological distance estimators \citep[e.g.,][]{Kowal68, Riess98, Perlmutter99, Riess11, Betoule14}.

Cosmological studies of SN Ia have always used some form of light-curve fitting \citep[e.g.,][]{Leibundgut91, Riess95, Riess96, Perlmutter97, Jha07, Guy07, Conley08, Burns11}, in order to interpolate their peak magnitudes and perform a K-correction from observed magnitudes to the rest-frame. These light-curve fitters traditionally consist of a (multi-colour) time-series light-curve template or SED built from a sample of well-observed, and often low-redshift, SNe Ia. However, although these fitters have worked well for cosmological applications, they have some limitations that a data-driven approach may not. 

Firstly, current template-based light-curve fitters are limited by the choice of light-curve parameters to be extracted and do not have the flexibility to easily extract further information from the SNe (for example, colour evolution) to understand their physics. Secondly, template-based fitters typically only work well where they are trained: typically on samples of \lq normal\rq\ SNe Ia with optical data, and excluding (for example) SN1991T- and SN1991bg-like events \citep{Filippenko92a, Filippenko92b, Ruiz-Lapuente93} or wider wavelength ranges such as the near-infrared (IR). Finally, alternative standardisations with different light-curve parameters can be explored: uncertainties due to the parametrizations used are currently an important factor in the systematic uncertainty budget of SN Ia cosmological analyses \citep[e.g.,][]{Abbott19}.

This paper is organised as follows. In Section~\ref{sec:pisco} we introduce and describe our new light-curve fitter. In Section~\ref{sec:test} and \ref{sec:salt2_comparison} we test our fitting code on samples of SNe Ia, both simulated and real. In Section~\ref{sec:lc_analysis} we then present an application of \pisco for SN Ia distance estimation, and explore alternative standardisations that \pisco can provide. A simple analysis of the SN Ia colour law derived from our light-curve fitter is shown in Section~\ref{sec:colour_law}, and we summarise in Section~\ref{sec:summary}. Throughout, we assume a flat $\Lambda$CDM cosmology with $H_0$ = 70\,km\,s$^{-1}$\,Mpc$^{-1}$, and $\Omega_{\rm M}$ = 0.3.


\section{PISCOLA}
\label{sec:pisco}

In this section, we introduce the light-curve fitter \pisco\ (Python for Intelligent Supernova-COsmology Light-curve Analysis). The code is open source\footnote{\url{https://github.com/temuller/piscola}} and written in \textsc{python}~3, a language widely used in the astronomy community. \pisco\ is intended to be simple and transparent in operation, allowing the user the opportunity to understand every step of the light-curve fitting and correction process. The full documentation can be found online\footnote{\url{https://piscola.readthedocs.io/en/latest/}}.

\begin{figure*}
	\includegraphics[width=\textwidth]{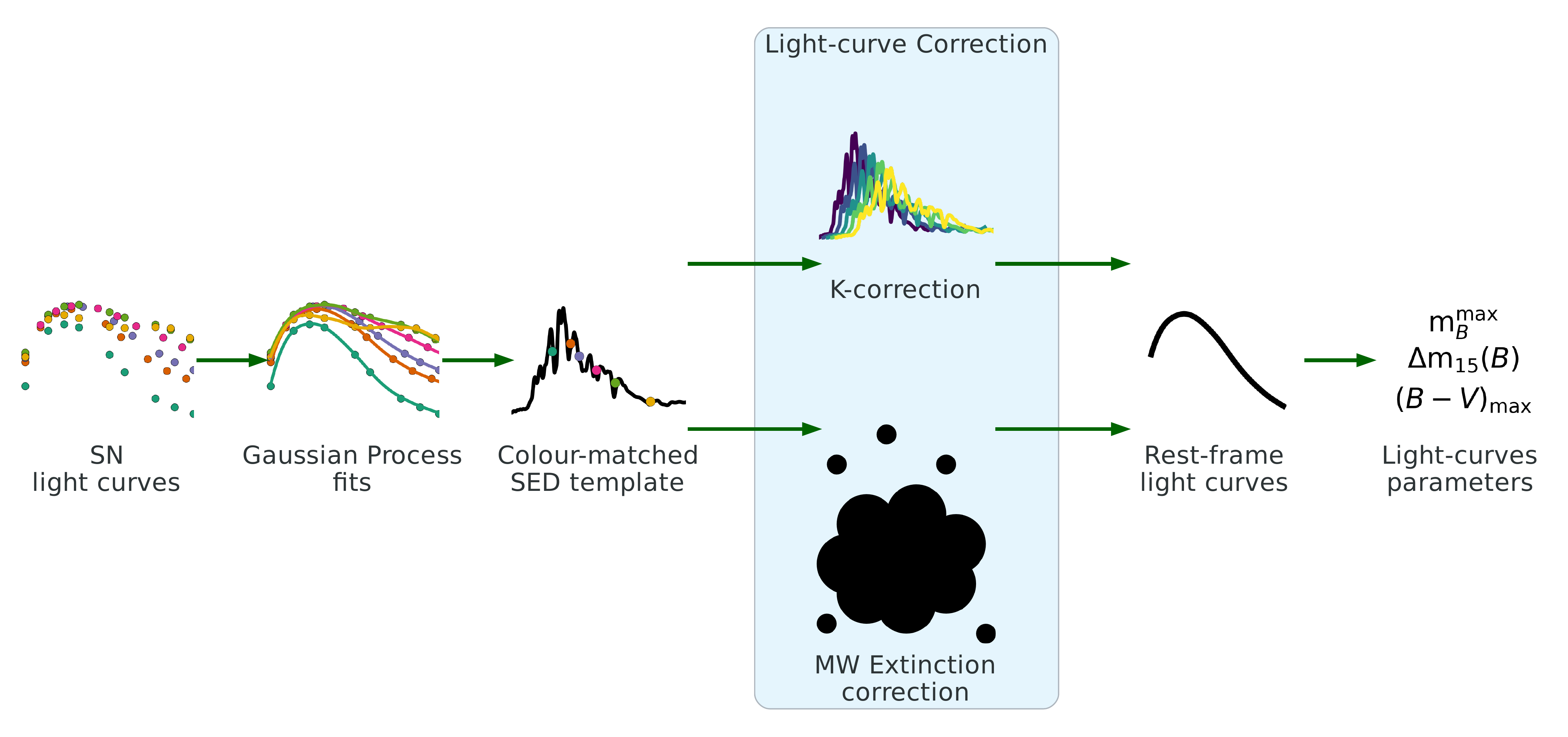}
    \caption{Flowchart of the main steps in the \pisco\ process. Gaussian processes are used to fit the SN light-curves as a function of wavelength and time (Section~\ref{subsec:lc_model}). An SED is then \lq warped\rq\ to match the observed SN colours, corrected for redshift ($K$-correction) and then Milky Way dust extinction (Section~\ref{subsec:sed_model}). Finally, the required rest-frame light curves are obtained and, optionally, light-curve parameters estimated from those light curves.}
    \label{fig:flowchart}
\end{figure*}

We introduce \pisco\ in the context of fitting SN Ia light curves and subsequently in SN Ia distance estimation, but stress that \pisco\ is flexible and can be used to fit the light curves of any optical transient. In Fig.~\ref{fig:flowchart}, we show the main steps in \pisco, described in the following sections.

\subsection{The light-curve model}
\label{subsec:lc_model}

\pisco\ uses Gaussian processes (GPs), a Bayesian method, to model the SN light curves by assuming that the data distributes as a multi-variate Gaussian distribution, which is the case for most astronomical observations. Unlike many other SN light curve fitters, there is no underlying SN template or model that is directly fit to the observed light curves; the observer-frame modelling is entirely data driven. This implies that our method needs better quality data than template-driven methods to reliably fit SNe Ia: it is not suited to poorly sampled data. In Section~\ref{sec:test}, we determine the quality of the data necessary for \pisco to produce reliable results. 

GP is an excellent tool for data regression, providing a more robust regression than polynomials as they can be undesirably global (splines are just a special case of GP regression). GPs have also been previously used in the context of SN and SN-like light curves, with excellent results \citep[e.g.,][]{Kim13, deJaeger17, Inserra18, Angus19, Pursiainen20, Wiseman20}.

A GP model is defined by a mean and a covariance function/kernel. There are different types of kernels (e.g., \lq squared exponential\rq, \lq rational quadratic\rq, \lq periodic\rq, the \lq Mat\'ern family\rq, etc.), each with its own set of hyperparameters (see Appendix~\ref{app:gp} for more details). \pisco makes use of \textsc{george} \citep{Ambikasaran16}, a GP implementation in \textsc{python} that allows the user to choose between several well-known kernels, together with the \textsc{scipy} and \textsc{lmfit} packages for the optimisation routines. Our code implements three kernels: squared exponential, \mtt and \mft. The Mat\'ern family is broadly used in different areas of research as it is effective at describing different physical processes \citep{Rasmussen06}, such as light curves, while the squared exponential kernel provides a smoother fit. We use different kernels at different points in our light-curve fitting process.

The fitting of the SN multi-colour observer-frame light curves is performed in two dimensions, as a function of time and wavelength. All three hyperparameters are optimised at this stage: length scales for the time and wavelength axes, and the variance (see Appendix~\ref{app:gp} for details about these hyperparameters). We fit in magnitude space as the logarithmic scale provides smoother and more accurate GP fits than using fluxes directly; when fitting in flux space, any large differences in flux between different filters can produce fits with under-estimated peaks (this happens for different choices of kernel).

There are two disadvantages of fitting in magnitude space. The first is that non-detections are not used as they cannot be represented correctly in magnitude space. This also means that data around peak luminosity will have more weight than data in the tails of the light curves, for measurements with the same uncertainty in flux space, during the regression. We note that it is the data around peak that interests us most in the light-curve fitting for the applications in this paper. A second disadvantage is that lower S/N observations have asymmetric errors, which are an issue for GP regression. However, these can be removed by masking out these observations (see Section~\ref{sec:test}), although this reduces the amount of useful data.

\begin{figure}
	\includegraphics[width=\columnwidth]{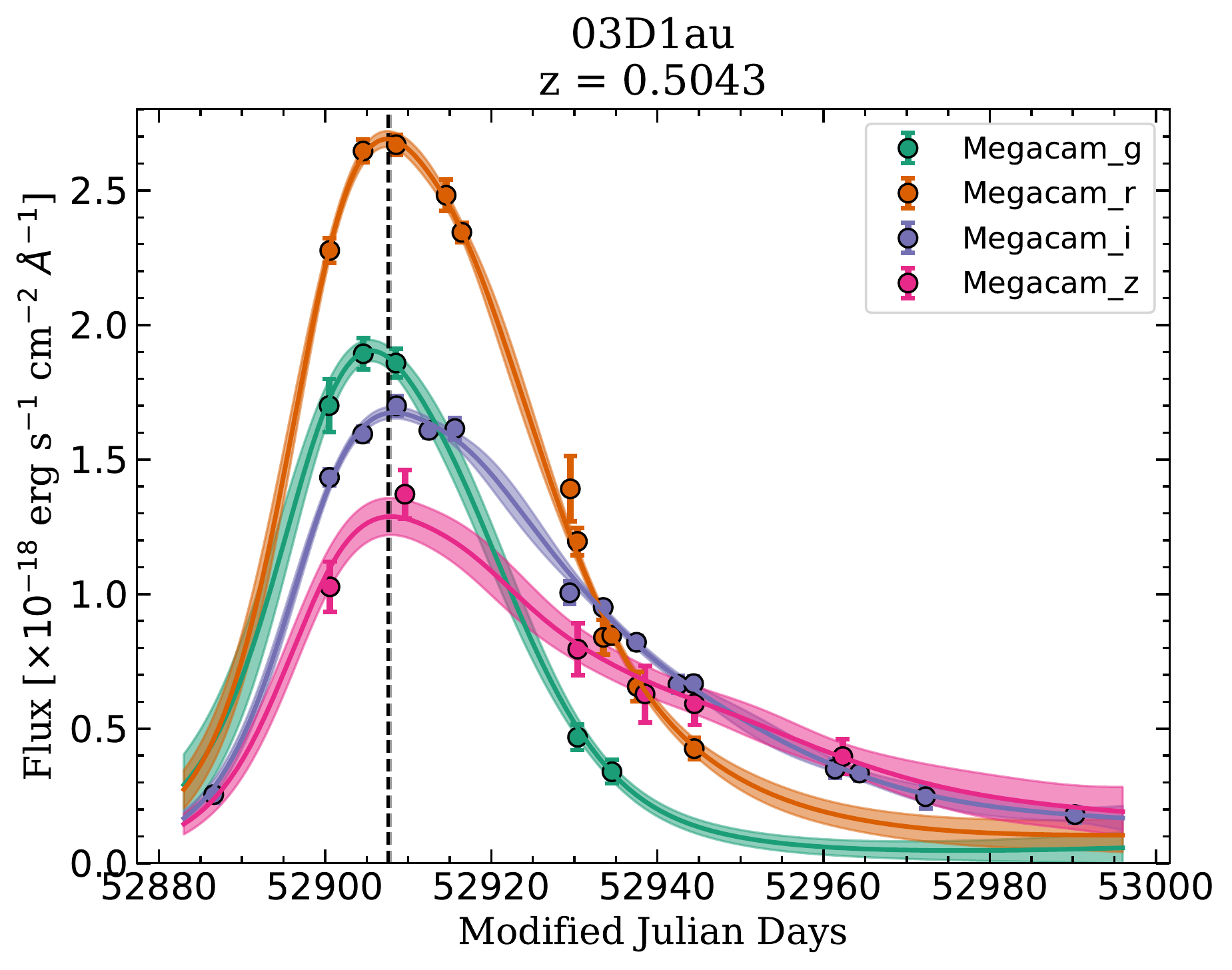}
    \caption{Two dimensional GP model of the observer-frame multi-colour light curve of the SN Ia SNLS-03D1au using a \mft\ kernel. The original data, from the Supernova Legacy Survey \citep{Astier06}, were observed in the Canada--France--Hawaii Telescope MegaCam $griz$ filters and are shown as circles with uncertainties. The solid lines show the mean of the GP model, and the shaded areas represent one standard deviation (1$\sigma$). The vertical dashed-black line marks the estimation of the time of rest-frame $B$-band maximum light.}
    \label{fig:lc_fits}
\end{figure}

The two-dimensional regression provides more robust results than fitting in one dimension at a time, i.e., flux as a function of time independently for each band, as it uses information from multiple bands to cover time gaps in different filters allowing an improved interpolation (see Appendix~\ref{app:1d_gp}). An example of the fitting applied to a high-redshift SN Ia $griz$ light curve is shown in Fig.~\ref{fig:lc_fits}.

A \mft kernel is the default option in \pisco as it produces flexible, yet smooth fits. In Fig.~\ref{fig:lc_fits}, the fit to the redder $z$-band is good, despite having larger uncertainties. In addition, although only the $i$-band has relatively good coverage of the tails of the light curve both pre- and post-peak, the two-dimensional GP model allows an informative extrapolation of the time axis of the $grz$-band light curves. For comparison, an example with a one-dimensional GP model is shown in Fig.~\ref{fig:lc_fits_1D} and described in Appendix~\ref{app:1d_gp}.

A further advantage of fitting in two dimensions is that more accurate estimations of the time of peak luminosity in a given rest-frame filter are obtained compared to 1D fits in individual observer-frame filters. The time of peak luminosity in a given band is estimated from the GP fit by calculating the time at which the derivative becomes zero at the effective wavelength of the desired band. For the case of the rest-frame $B$ band, we denote the time of peak luminosity as parameter \tmax. This is important in many applications, but particularly in the light-curve correction processes for SNe Ia in cosmology. 

Moreover, \pisco can produce excellent results with relatively well-sampled SNe Ia, such as those at low redshift ($z \lesssim$0.1), as it is able to fit bands outside the useful ranges of other light-curve fitters. An example is shown in Appendix~\ref{app:lc_fits}. This is of great importance in many areas, for example in the estimation of the Hubble constant, measurement of peculiar velocities in the local universe, and the extraction of astrophysics from the SNe Ia light curves, to name just a few applications.


\subsection{The spectral energy distribution model}
\label{subsec:sed_model}

Although no light-curve template is used in \pisco, a (time-dependent) SED is still required to K-correct the observer-frame light curves to rest-frame filters. Many such templates are available for different SN types \citep[e.g.,][]{Nugent02, Blondin07, Hsiao07, Rodney09, Liu14, Vincenzi19}. We colour-adjust (or \lq mangle\rq) the SED as a smooth function of wavelength so that it reproduces the colours of the observed light curves at each epoch, as estimated by the GP model \citep[see also][]{Hsiao07, Conley08}. The result is a colour-matched SED, i.e., a SN time-series SED which reproduces the observed light curves of the SN being fit.

The colour-matching is performed by multiplying the SED by a wavelength-dependent function at the desired phases (with respect to \tmax), after which the SED reproduces the observed light-curve when integrated through the observed filters. This wavelength-dependent function is often represented by a spline in the literature, but \pisco makes use of GPs with (by default) a squared exponential kernel, resulting in a smooth function and a natural way of propagating uncertainties in the process. The steps for developing the colour-matching function for each SN are:

\begin{enumerate}
\item An observer-frame SED is constructed from the target's redshift;
\item Milky Way (MW) dust extinction is applied to the SED;
\item The observed filters are used to calculate model fluxes from the SED;
\item The ratios between the actual fluxes of the observed SN and the model fluxes from the SED are calculated together with their respective effective wavelengths;
\item A wavelength-dependent multiplicative function is calculated by modelling the flux ratios with GPs using an squared exponential kernel. 
\end{enumerate}

\begin{figure}
	\includegraphics[width=\columnwidth]{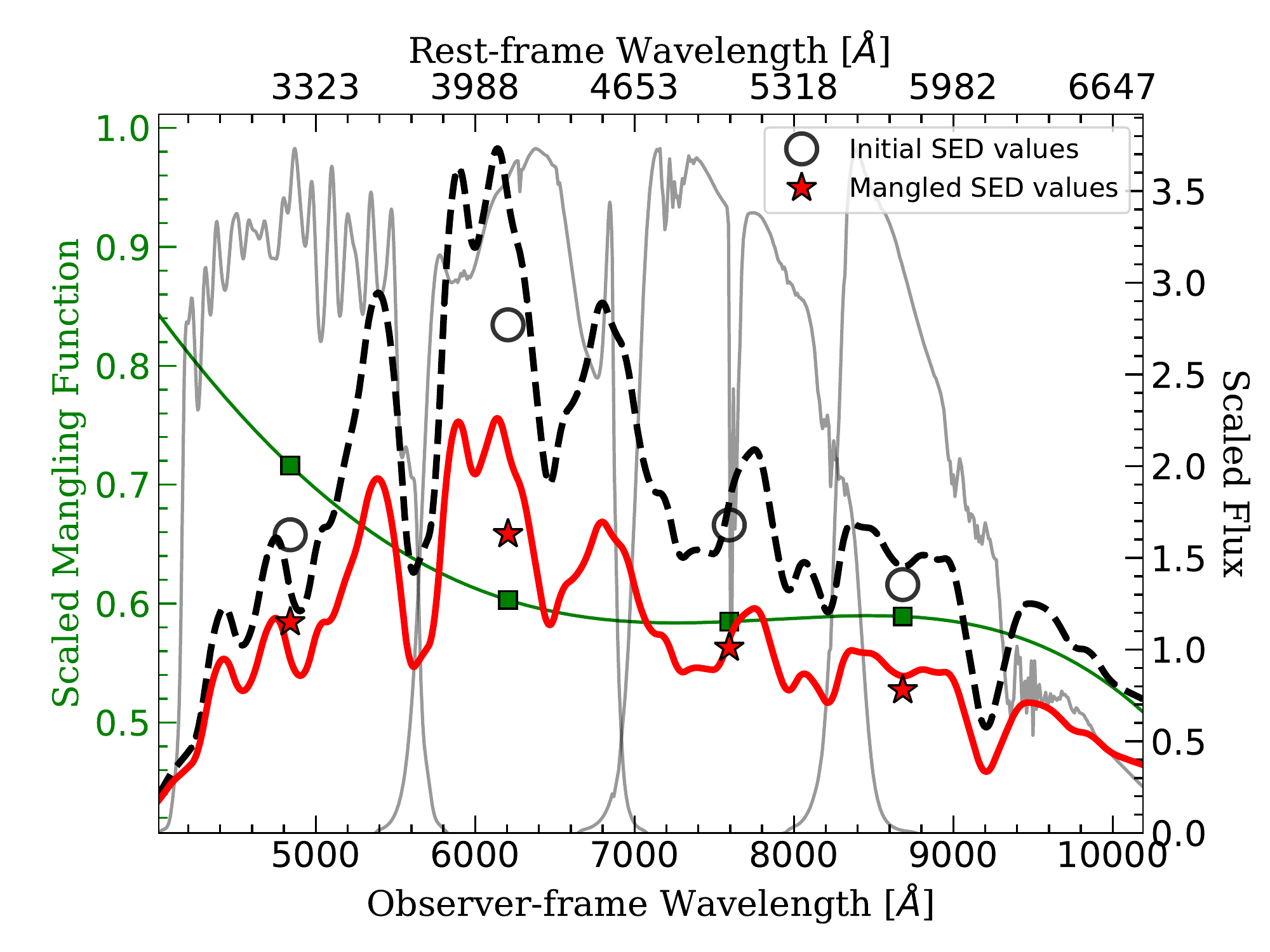}
	\caption{Mangling function (solid green line) for SNLS-03D1au at \tmax. The green squares represent the flux ratios of the different bands centered at their respective effective wavelengths. The initial SED (dashed black line) is compared to the SED multiplied by the mangling function (solid red line). The initial (black open circles) and mangled (red stars) SED fluxes are also shown. The transmission functions of the filters used are plotted in grey. The scaling is arbitrary for visualisation.}
    \label{fig:mangling}
\end{figure}

The SED template is multiplied by the multiplicative function to produce a colour-matched SED. An example of a typical multiplicative function and SED template is shown in Fig.~\ref{fig:mangling}.

As a final step, the colour-matched SED is corrected for MW extinction and shifted to the rest-frame to obtain a final SED model. \pisco corrects for MW dust extinction using the \citet{Schlafly11} dust maps and a \citet{Fitzpatrick99} extinction law as default. Other implementations of the dust maps \citep{Schlegel98} and extinction laws are also available. The \textsc{sfdmap} and \textsc{extinction} \textsc{python} packages are used to perform the corrections. \pisco does not estimate, or correct for, any other sources of extinction. This procedure is repeated for as many phases as desired for each SN, depending on the data coverage.

The product of this process is a rest-frame time-series SN SED that reproduces the observer-frame light curves of the SN being modelled. From this, rest-frame light curves in any required band can be calculated within the wavelength and phase limitations of the input data (together with any desired light-curve parameters based on those rest-frame light curves). This process is agnostic to the physical type of transient being fit, under the assumption that a physically relevant time series SED is used in the colour-matching process.

We note one subtlety: the procedure described above depends on the initial estimate of \tmax. This initial estimate can be improved using the rest-frame light curves. If the initial and improved estimates of \tmax are not consistent, the light-curve fitting process can be repeated with the new estimate until a satisfactory convergence is reached.

\begin{figure}
	\includegraphics[width=\columnwidth]{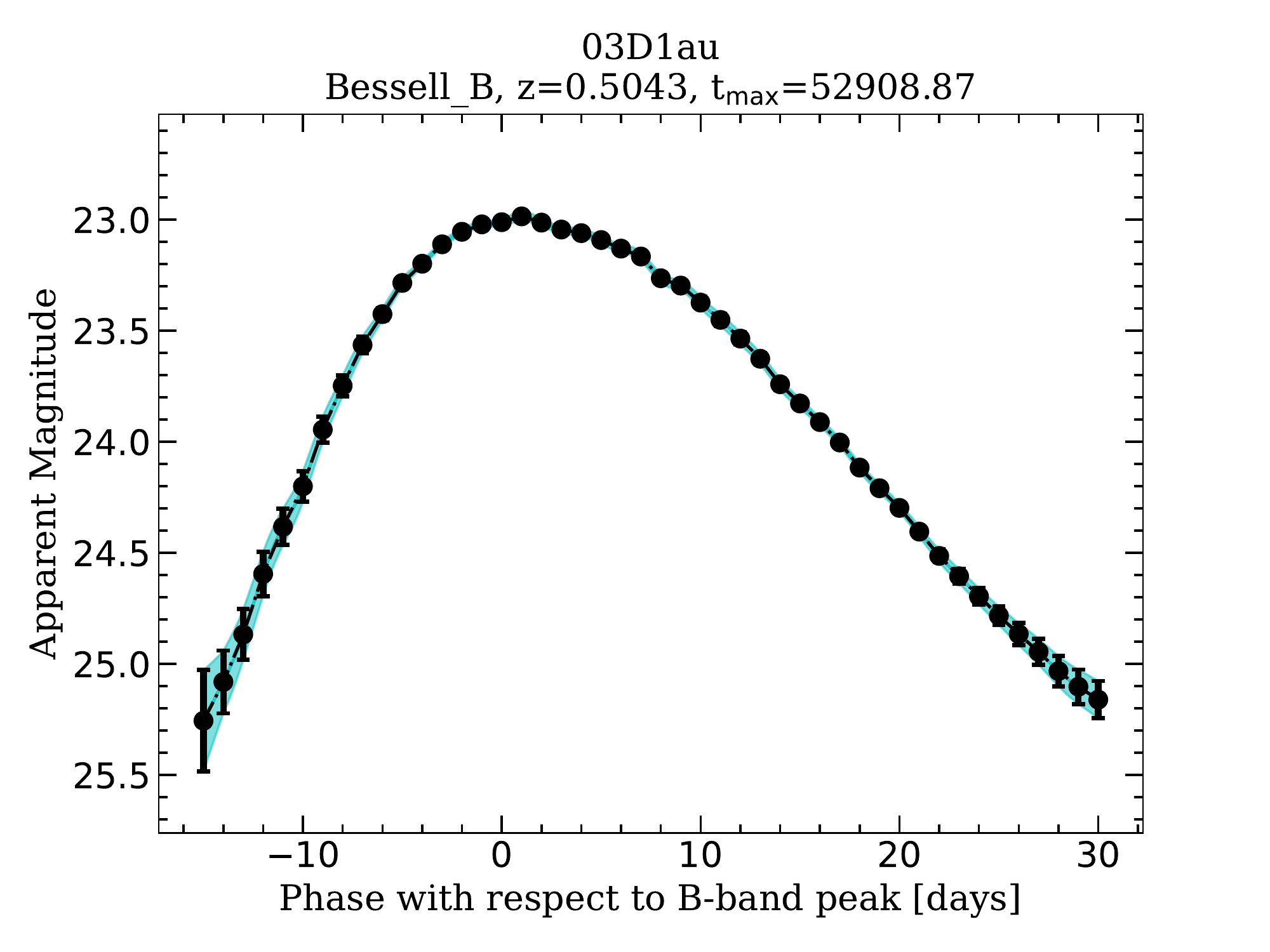}
    \caption{Rest-frame $B$-band light curve of SNLS-03D1au. In this case, observations with Megacam-$griz$ filters were used to construct the $B$-band light curve.}
    \label{fig:Bband}
\end{figure}

In the remainder of this section, and to asses the performance of \pisco, we discuss the specific use of \pisco\ to the fitting of SN Ia light curves. Here we focus on traditional SN Ia rest-frame light-curve parameters: the rest-frame $B$-band peak apparent magnitude (\mb), the decline in magnitudes in the 15 days following \tmax (\dm), and the $B-V$ colour at \tmax (\colour). In Fig.~\ref{fig:Bband}, we show an example of a resulting rest-frame $B$-band light curve of a SN Ia.

\subsection{Application to SNe Ia} 
\label{subsec:calibration}

Several aspects of the light-curve fitting process can be affected by choices in the analysis, including the MW extinction law, the SED time-series template and the exact rest-frame filters. In Section~\ref{sec:salt2_comparison}, when we will compare \pisco with results from the SALT2 SN Ia light-curve fitter to establish its performance, an accurate comparison will require these choices to be consistent.

SALT2 is an empirical SN Ia model, used in several major cosmological analyses \citep[][]{Betoule14, Scolnic18, Abbott19}, trained on a spectrophotometric sample of SNe Ia. It describes the flux of a SN Ia with the following functional form,
\begin{equation}
\begin{aligned}
F(S N, p, \lambda)=x_{0} & \times\left[M_{0}(p, \lambda)+x_{1} M_{1}(p, \lambda)+\ldots\right] \\
& \times \exp [c \times C L(\lambda)],
\end{aligned}
\label{eq:salt2}
\end{equation}
where $p$ is the rest-frame time with respect to $B$-band peak, $\lambda$ is the rest-frame wavelength, $M_0$ is the average spectral sequence, $M_1$ is an additional component that describes further variability, and $CL$ represents the average colour variation law of a SN Ia (Section~\ref{sec:colour_law}). The contribution of higher-order components is less significant and not used. The terms $x_0$, $x_1$ and $c$ are the SALT2 light-curve parameters, where $-2.5\log_{10}(x_0) + 10.635 = $\mb, $x_1$ is a measurement of the stretch of the light curve, and $c$ is the colour of the SN.

We ensure our choices follow \citet{Scolnic18} as far as possible:
\begin{itemize}
    \item We adopt the extinction law from \citet{Fitzpatrick99};
    \item The SN Ia SED time-series template is the $M_0$ component from the SALT2 model (\texttt{salt2\_template\_0.dat}). We do not include the SALT2 $x_1$-dependent component in the SED as \pisco does not measure this parameter (i.e., a template with $x_1=c=0.0$ is used);
    \item We use the same filter transmission functions for the observed light curves as \citet{Scolnic18};
    \item We use the same photometric calibration system for magnitude systems from different surveys (e.g., AB using BD $+17^{\circ}4708$ \citep{Bohlin04b} as primary standard);
    \item We use the Bessell filters \citep[][]{Bessell90}, shifted to match \citet{Landolt92} observations, as included in SALT2 \citep[see][and Appendix A of \citealt{Conley11}]{Betoule14}, to estimate the rest-frame light-curve parameters.
\end{itemize}

In principle, the use of the $M_0$ SED but neglecting the spectral differences introduced by $x_1$, will lead to small differences in the $K$-corrections when compared to SALT2. For example, when including the $x_1$-dependent component (assuming the $x_1$ values from \citealt{Scolnic18}), the differences in the rest-frame $B$-band light curve are on average $\lesssim 0.01$\,mag around peak and $\lesssim 0.03$\,mag at earlier or later phases. This sets a natural limit to the accuracy of these tests.


\section{Tests with Simulations of SNe Ia}
\label{sec:test}

We test \pisco with extensive simulations using the SuperNova ANAlysis software \citep[\textsc{snana};][]{Kessler09b}, version \texttt{v10\_75c}. We focus our tests on the effects of cadence, the mean time between consecutive observations of an object in the same filter, and observational uncertainties or, equivalently, signal-to-noise ratio (S/N).

\textsc{snana} simulations simulate transient surveys and produce light curves of simulated transient events, accounting for the survey observing pattern, limiting magnitudes, pointing on the sky, and so forth. For testing \pisco, we simulate Pantheon-like SN Ia samples \citep{Scolnic18}, representing one of the most comprehensive compilations of SNe Ia. Pantheon comprises many different SN Ia surveys: the Harvard-Smithsonian Center for Astrophysics (CfA) surveys 1--4 \citep[][]{Riess99, Jha06, Hicken09a, Hicken09b, Hicken12}, the Carnegie Supernova Project \citep[CSP;][]{Contreras10, Folatelli10, Stritzinger11}, the Sloan Digital Sky Survey SN Survey \citep[SDSS;][]{Frieman08, Kessler09a, Sollerman09, Sako18}, the Supernova Legacy Survey \citep[SNLS;][]{Astier06, Guy10}, the Panoramic Survey Telescope and Rapid Response System 1 (Pan-STARRS1 or PS1) Medium Deep Survey \citep{Rest14, Scolnic14} and different \textit{Hubble Space Telescope} (HST) surveys, including: the Supernova Cosmology Project \citep[SCP;][]{Suzuki12}, the Great Observatories Origins Deep Survey \citep[GOODS;][]{Riess07}, and the Cosmic Assembly Near-infrared Deep Extragalactic Legacy Survey + Cluster Lensing And Supernova Survey with Hubble \citep[CANDELS+CLASH;][]{Graur14, Rodney14, Riess18}.

We denote the compilation of the CfA1--4 and CSP surveys the \lq low-$z$\rq\ sample (or survey) as they only contain SNe Ia at $z <$0.1. For a more detailed description of the surveys, see \citet{Scolnic18} or their respective references. We use the SALT2 light-curve parameters for these SNe Ia from \citet{Scolnic18}\footnote{\url{https://archive.stsci.edu/prepds/ps1cosmo/}}.

All simulations have the same main characteristics. We restrict our simulations to the ground-based samples (low-$z$, SDSS, SNLS and PS1) as \pisco does not perform well with the less well-sampled light curves that are typical from the \textit{HST} surveys. Simulated SNe Ia are based on the SALT2 model from \citet[][version 2.4]{Betoule14}, also used by \citet{Scolnic18}. The SNe are simulated by selecting random $x_1$ and $c$ values drawn from the distributions of \citet[][]{Scolnic16}, which are survey dependent based on several factors such as Malmquist bias; for more details, see \citet{Scolnic16}. Approximately 500 SNe are simulated for each SN sample (low-$z$, SDSS, SNLS and PS1) and for each individual test. In other words, each simulated Pantheon-like sample contains a total of $\sim$2000 SNe Ia. We discuss this choice in Section~\ref{subsec:fitting_time}. 

The comparison between the fits and the simulations are based on \mb, i.e., we focus on the residual between the simulated \mb value and the estimated value obtained with \pisco. In this paper, we will refer to this as $\Delta$\mb $\equiv m^{\rm max}_{B, \,\text{PISCOLA}}-m^{\rm max}_{B, \,\text{simulations}}$. We fit the SNe in an automated way with the default GP kernels and over $-$15 to $+$30 days with respect to \tmaxB (default phases). In addition, we require every SN to have a sufficient coverage of the peak: at least one data point in any band over $-$7 and 0 days, one from 0 to $+$7 days, and one from $-$3.5 and $+$3.5 days (this final constraint may overlap with one or both previous constraints). We also mask out observations with a S/N $\leq5$ to prevent poor-quality fits and asymmetric errors. We note that applying cuts in S/N does not bias the estimation of \mb or the observed peaks in the fitted light curves (this was tested on real data and simulation from the following sections), although it could possibly bias the fits at earlier or later epochs. However, for the analysis in this work, we are mainly concerned about the data around peak. Note that the cuts applied are different, but more stringent than those applied in other analyses using SALT2 \citep[e.g.,][]{Scolnic18}, discarding more SNe.

These constraints reduce the sample of SNe, but they ensure a relatively good quality light curve data required by GP interpolation (to be tested in the following sections). The chosen values are somewhat arbitrary, but small changes do not significantly alter the results. We note that, although template-driven fitters can fit SNe with different S/N, they may introduce biases as, in the case of low S/N, the fitters will mainly retrieve the light-curve parameters of the templates being used.


\subsection{SN Ia Pantheon sample simulations}
\label{subsec:pantheon_sim}

\begin{figure*}
	\includegraphics[width=\textwidth]{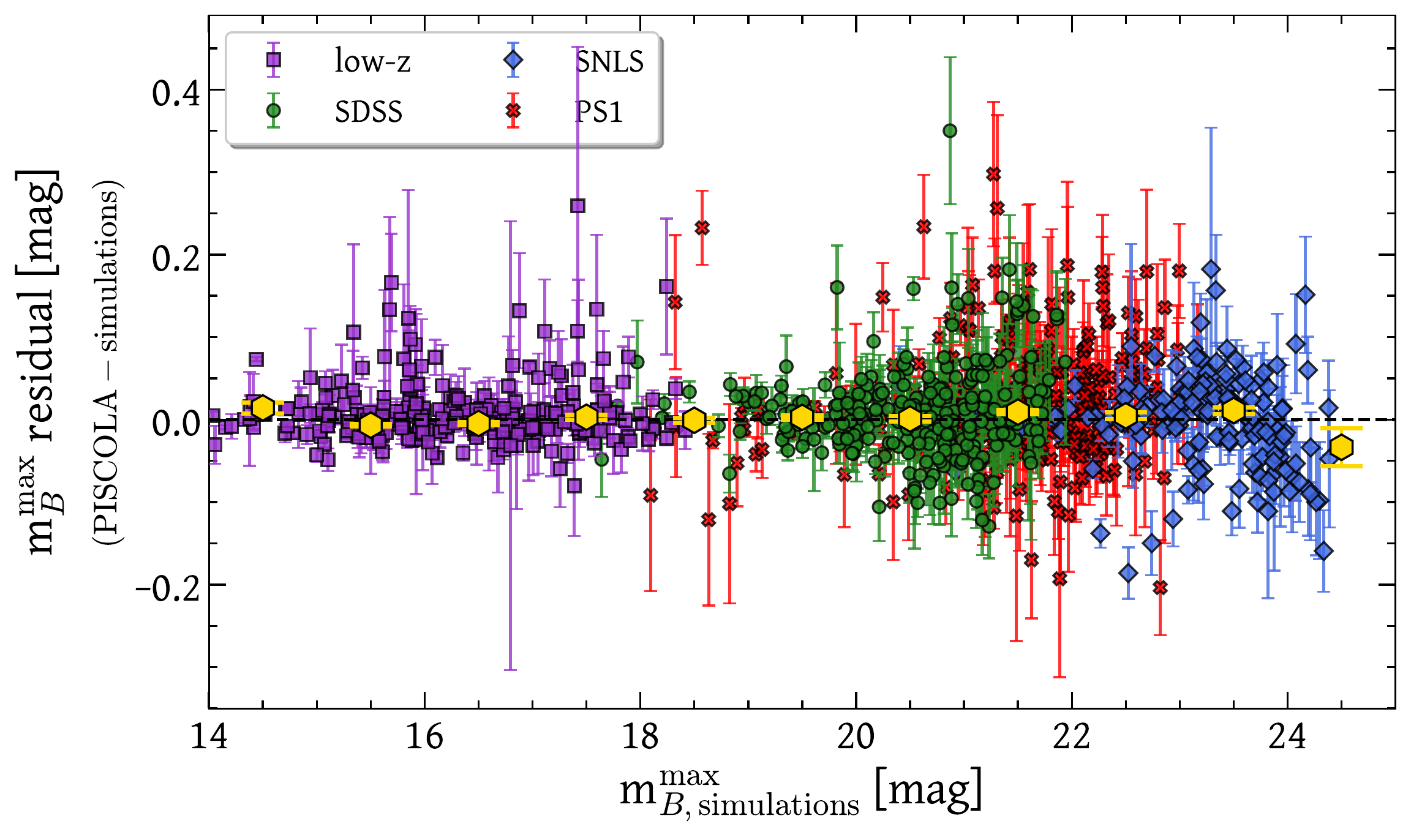}
    \caption{$\Delta$\mb $\equiv m^{\rm max}_{B, \,\text{PISCOLA}}-m^{\rm max}_{B, \,\text{simulations}}$ for a Pantheon-like simulation. Each point represents a single simulated SN, colour-coded according to the legend. The uncertainties are 1$\sigma$ and are taken from the \pisco fits. The golden hexagons represent the weighted mean in bins of 1\,mag with their respective uncertainties (1$\sigma$).}
    \label{fig:mb_comparison_pantheon_sim}
\end{figure*}

The first step is to test \pisco with a Pantheon-like simulation. The comparison between the \pisco-measured and simulated \mb values is shown in Fig.~\ref{fig:mb_comparison_pantheon_sim}.

We discard some SNe at different stages of the fitting process for several reasons:

\begin{enumerate}
    \item The observer-frame wavelength coverage does not cover the rest-frame $B$-band at the redshift of the SN (e.g., SNe at high redshift);
    \item The temporal coverage does not allow an initial estimation of \tmax (e.g., many low-$z$ do not have data prior to the $B$-band peak);
    \item \pisco is unable to estimate an accurate $B$-band light curve. This happens when \pisco does not converge to an estimation of \tmax; e.g., several low-$z$ SNe have their $B$-band peak only partially covered, and \tmax can fall close to the limits of the coverage, producing a failure to estimate a new peak; 
    \item The light-curve does not satisfy the additional constraints on peak coverage (Section~\ref{sec:test}), mainly due to a combination of low cadence and/or masking of low S/N data;
    \item Visual inspection reveals poor \pisco fits, mainly caused by remaining poor data quality.
\end{enumerate}

The first three reasons are automated by \pisco. Discarding SNe after visual inspection is only feasible due to the relatively low number of SNe. This is a limitation, and for future implementations we aim to automate it with a statistically-motivated metric, which is the aim of this set of tests. We note that future surveys (e.g., LSST), will produce higher quality observations with lower uncertainties and better cadences.

We successfully fitted (obtained \mb) $\sim$50 per cent of the SNe. This is a low number compared to other light-curve fitters, but highlights the relatively poor data quality of some historical surveys that SNANA simulates. Stages ($ii$) and ($iv$) are the main causes of discarding SNe (see Section~\ref{sec:salt2_comparison}). We note that these discarding reasons do not strongly depend on redshift but mainly on the observing strategy of each survey.

In Table~\ref{tab:pantheon_sim}, we show the values of $\Delta$\mb. There are no significant (i.e., $<$3$\sigma$) deviations from $\Delta$\mb $=$ 0.0\,mag for all surveys, except for PS1 at 3.5$\sigma$ significance but small ($<0.01$\,mag) deviation. This is mainly caused by a few SNe with underestimated \mb and small uncertainty. Additionally, Fig.~\ref{fig:mb_comparison_pantheon_sim} shows that some low-$z$ SNe have underestimated \mb values, while some SNLS SNe have the opposite. This is most likely due to a combination of cadence and S/N and is further investigated in the following section. Generally, we conclude that \pisco is successful at retrieving accurate \mb values at $<0.01$\,mag.

\begin{center}
\begin{table}
\centering

\caption{Weighted mean, uncertainty on the weighted mean and weighted standard deviation of $\Delta$\mb for a Pantheon-like simulation (Sec.~\ref{subsec:pantheon_sim}).}

\begin{tabular}{cccc}
  &   &  error on the  &  weighted\\
   survey & weighted mean &  weighted mean &   standard deviation \\
     & (mmag) & (mmag) & (mmag) \\
\hline
low-$z$       & 1        &    1            & 24                                         \\
SDSS          & 4       & 2                  & 34                                         \\
SNLS          & 3        & 2              & 29                                         \\
PS1           & 7       & 2                    & 35                                        \\


\end{tabular}
\label{tab:pantheon_sim}
\end{table}
\end{center}


\subsection{Effect of observational cadence}
\label{subsec:testing_cadence}

\begin{figure}
	\includegraphics[width=\columnwidth]{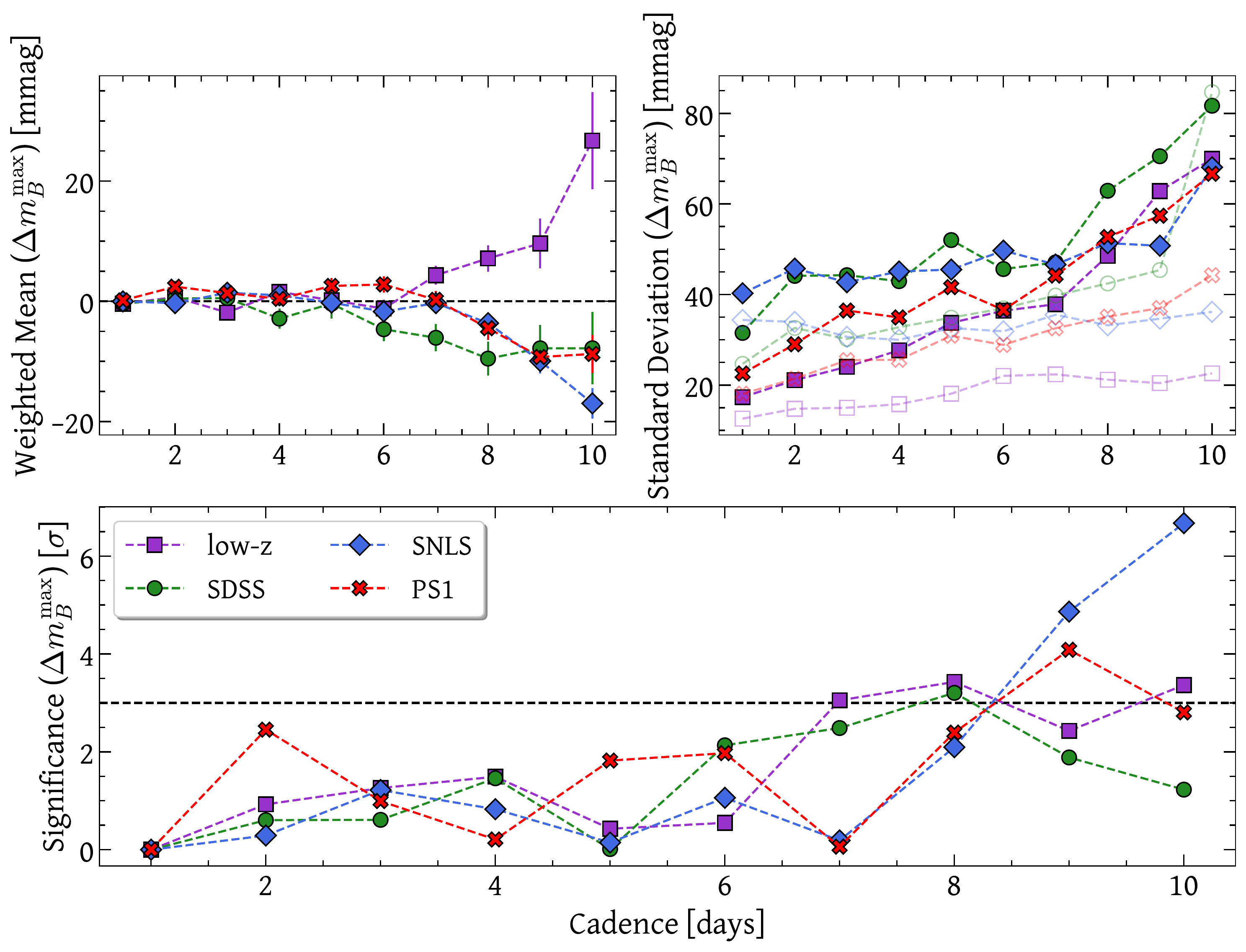}
    \caption{Weighted mean with uncertainty on the weighted mean (\textbf{top left} panel) and standard deviation (\textbf{top right} panel) of $\Delta$\mb for simulations with cadences between 1 and 10 days. The standard deviation of $\Delta$\mb for SALT2-measured \mb values, i.e., $m^{\rm max}_{B, \,\text{SALT2}}- m^{\rm max}_{B, \,\text{simulations}}$, are also shown for comparison (transparent symbols in the \textbf{top right} panel). Measured significant deviations of $\Delta$\mb $=$ 0.0\,mag in units of $\sigma$ are also shown (\textbf{bottom} panel). The horizontal line in the bottom panel marks a significance of 3$\sigma$.}
    \label{fig:mb_offset_evolution}
\end{figure}

As \pisco is  a data-driven fitting method, the cadence is important as the GP model has no prior for the true shape of the SN light curves. We therefore simulated a set of Pantheon-like samples with cadences between 1 and 10 days in steps of 1 day, and took the 1-day cadence simulation as our benchmark. Observations are equally spaced for all bands (e.g., simultaneous/same-day $ugriz$-bands observations every $x$ days) and with random characteristics drawn from survey-dependent distributions. We then estimate the reliability of \pisco as the cadence becomes poorer. The results are shown in Fig.~\ref{fig:mb_offset_evolution}. 

We successfully fitted (obtained \mb) $\sim$75--85 per cent of the SNe in each of the 1- to 7-day cadence simulations, and $\sim$60--75 per cent for the 8- to 10-day cadence simulations (SALT2 fits $\gtrsim$98 per cent of the simulated SNe in all cases). The unsuccessful SNe are mainly due to low-S/N light curves in the SDSS, SNLS and PS1 simulated samples and $B$-band peak not well covered (e.g., data only after peak) in the case of low-$z$ objects. SNe were discarded for the same reasons as in Section~\ref{subsec:pantheon_sim}. At relatively high cadence ($\lesssim$7\,d) \pisco is accurate when estimating \mb for the different surveys (i.e., $\lesssim$ 3$\sigma$ deviations from $\Delta$\mb $=$ 0.0\,mag; see bottom panel of Fig.~\ref{fig:mb_offset_evolution}). However, the accuracy is improved for SNLS and PS1 compared to low-$z$ and SDSS samples, given their superior S/N and/or rest-frame cadence.

As redshift increases, the rest-frame cadence increases for fixed observer-frame cadence. This gives improved light-curve coverage, but with a trade-off of S/N. We examined $\Delta$\mb versus redshift, but see no clear trend. The surveys at different redshifts also generally perform similarly as a function of cadence. We find that \pisco accurately retrieves \mb ($\lesssim$ 3$\sigma$ deviations from $\Delta$\mb $=$ 0.0\,mag) for cadences similar to those of most high-redshift transient surveys, which have a typical cadence of about 7 days in the observer frame (e.g., Dark Energy Survey).

However when the cadence is $\gtrsim$7\,days, the performance of \pisco on the low-$z$ simulated sample is quantitatively different to the performance on the higher-redshift simulations. At low-$z$, \pisco under-predicts \mb ($\Delta$\mb is positive). This is because the peaks of the light curves are smoothed out: the GP model has no information about their true shape and thus does not recover the peak given a lack of information. In the other simulations (SDSS, SNLS and PS1), the opposite trend is seen: $\Delta$\mb is systematically negative at low cadences. This is due to the lower S/N in these data, despite the higher rest-frame cadence.

SNe with the lowest S/N, which would normally produce slightly under-predicted \mb values (see Fig.~\ref{fig:cad_and_unc_effect}), are not successfully fit, and therefore, not included in the comparison, producing the effect of an apparent over-prediction of \mb for these samples instead of just an increase in scatter. The masking of low-S/N data in part produces a similar effect. This explains some of the discrepancies seen for some SNLS SNe in Section~\ref{subsec:pantheon_sim}. SNe with high S/N, but relatively low cadences, can have their \mb values slightly over-predicted due to measurements with low uncertainty at each side of the light-curve peak, which can produces sharper peaks (top left panel of Fig.~\ref{fig:cad_and_unc_effect}). However, this is not the general case (see the top left panel of Fig.~\ref{fig:mb_offset_evolution}) as it very much depends on the time of the observations with respect to the actual light-curve peak and the apparent luminosity of the SN.

These effects can be reduced by tightening the constraints of the observations around peak, although this would in turn reduce the number of usable SNe. In practice, real observations are not evenly spaced, so these tests just provide a more general idea of how well \pisco performs.


\subsection{Effect of observational uncertainties}
\label{subsec:testing_obs_errors}

The GP interpolation depends on the uncertainties of the observations. Thus, we tested the effect of observational uncertainties by simulating two Pantheon-like samples with half and twice the observational uncertainties ($\sigma_{\rm obs}$) of the data in the original Pantheon sample, with a fixed cadence of 7 days.

The weighted mean and weighted standard deviation of $\Delta$\mb for the 7-day cadence simulations with original, half and twice the observational uncertainties are shown in Table~\ref{tab:sim_results}. From this comparison, we can see that a deviation from $\Delta\mb=0.0$\,mag (at $\sim3\sigma$) for the SDSS sample is persistent for different values of $\sigma_{\rm obs}$, which may be caused by a generally low S/N despite the S/N$\geq5$ selection used. For the low-$z$ sample, a significant (3.5$\sigma$) deviation from $\Delta\mb=0.0$\,mag (\pisco under-predicts \mb) is only observed for half $\sigma_{\rm obs}$ given the relatively low cadence, despite the high S/N, as discussed at the end of Section~\ref{subsec:testing_cadence}. In the case of SNLS, a significant deviation (4.5$\sigma$) in the estimation of \mb is only seen for twice $\sigma_{\rm obs}$, while for PS1 no significant deviations are observed.

Despite having some disagreement in the estimation of \mb, these are all $<0.01$\,mag: \pisco is in good agreement with established light-curve fitters. However, the characteristics of surveys like SNLS and, especially, PS1, are ideal for fitting SNe with \pisco as they have relatively high S/N and the high-$z$ observations allow a good light-curve coverage due to the relatively high rest-frame cadence. Surveys with the characteristics of the low-$z$ sample require high cadences ($\lesssim$6 days). Future surveys, such as the LSST, will produce high-quality data with good S/N and cadence \citep[e.g.,][]{Lochner18, Scolnic18b}, overcoming some of the limitations found in this work, thus allowing \pisco to produce reliable fits.

These comparisons help validate our code. However, we note the caveat that the simulations are based on the SALT2 model, which differs from other light-curve fitters. Nonetheless, the results of the tests performed throughout this section are promising as they help establish the reliability of \pisco. The level of discrepancies uncovered in this section would only be important in the use of SNe Ia as cosmological probes; in other astrophysical applications the level of disagreement is not significant.

\begin{center}
\begin{table}
\centering

\caption{Weighted mean, uncertainty on the weighted mean and weighted standard deviation of $\Delta$\mb for the 7-day cadence simulations with initial, half and twice the observational uncertainties ($\sigma_{\rm obs}$; see Sec.~\ref{sec:test}).}

\begin{tabular}{cccc}
       &  & error on the & weighted   \\
    survey  & weighted mean & weighted mean & standard deviation\\
      & (mmag) &  (mmag) &  (mmag) \\
\hline
\multicolumn{4}{c}{\textbf{Initial simulation}}\\
low-$z$       & 4             &  2              & 26                                         \\
SDSS          & -6    &  2                       & 39                                         \\
SNLS          & 0    &  2                      & 29                                         \\
PS1           & 0      &  1                     & 30                                        \\
\hline
\multicolumn{4}{c}{\textbf{half $\boldsymbol{\sigma_{\rm obs}}$ simulation}}\\
low-$z$       & 7       &  2                    & 26                                         \\
SDSS          & -8      &  2                     & 34                                         \\
SNLS          & 4      &  2                     & 29                                         \\
PS1           & 1      &  1                     & 29                                         \\
\hline
\multicolumn{4}{c}{\textbf{twice $\boldsymbol{\sigma_{\rm obs}}$ simulation}}\\
low-$z$       & 4     &  2                      & 28                                         \\
SDSS          & -8          &  3                 & 45                                         \\
SNLS          & -9        &  2                   & 36                                         \\
PS1           & -4        &  2                   & 36 \\


\end{tabular}
\label{tab:sim_results}
\end{table}

\end{center}


\subsection{Computational considerations}
\label{subsec:fitting_time}

\pisco is a computationally intensive light-curve fitter. Fitting the light curves is fast (of the order of seconds); however, calculating the mangling function can take longer (of the order of minutes for the default phase between $-$15 and 30 days with respect to \tmax). This is because a fixed GP model is used inside a minimization routine to calculate the mangling function. The length-scale of the GP model is fixed to a value of 20 to produce a smooth function, however, the ratios between the observed and model fluxes (see Section~\ref{subsec:sed_model}) are treated as parameters for the minimisation routine. We note that the wavelength axis is divided by 1000 before the minimization routine, and before setting the GP length-scale, to ensure better results by avoiding large numbers, as these are not always properly handled by the routine used. The results are then re-scaled by 1000.

The large covariance between the different bands, given by the \se kernel used (see Fig.~\ref{fig:covariance_kernels}), and the precision required, make the calculation of the mangling function a slow process as the minimisation routine takes longer to converge. Additionally, if the whole light-curve correction process needs to be repeated with an improved estimation of \tmax (Section~\ref{subsec:sed_model}), the time increases further.

As a result, the number of SNe that can be fit in a \lq reasonable\rq\ amount of time is limited to a few thousands or tens of thousands. For instance, fitting 1000 simulated SNe Ia takes at least $\sim$2000 minutes ($\sim$33\,hours) on a single CPU core, depending on various factors. This motivated our choice of $\sim$2000 simulated SNe per test, allowing us to fit objects in a reasonable amount of time.


\section{Comparison with SALT2}
\label{sec:salt2_comparison}

We next test \pisco on real data, using the 1022 SNe Ia from the Pantheon sample (excluding HST objects) for these tests. Unlike the simulations in the previous section, we have no \lq ground truth\rq\ for these tests, so instead we compare the outputs of \pisco against those of SALT2 using the light-curve parameter estimations as given in \citet{Scolnic18}. We fit the SNe in the same way as we described in the previous section (using default \pisco parameters) and applied the same constraints as for the simulations (see Section~\ref{sec:test}). SNe were discarded at different stages of the fitting process following Section~\ref{subsec:pantheon_sim}.

Of the 1022 initial SNe Ia, we obtain successful fits, with \mb values, for 620, of which 413 have a \colour estimate. Table~\ref{tab:sample_cuts} shows a summary of the discarded SNe. The percentage of successfully fitted SNe ($\sim$60 per cent) is larger than for the Pantheon-like simulation in Section~\ref{subsec:pantheon_sim} ($\sim$50 per cent), which is explained by the different relative numbers of SNe for the different surveys: in the simulations, all surveys have approximately the same number of SNe.

The outputs from different light-curve fitters are difficult to compare on a SN-by-SN basis due to, for example, different internal calibrations in the fitters producing offsets in some parameters. These are not important for cosmology as long as each code is self-consistent. In this section, we focus on the comparisons of \mb and \colour which are the most directly comparable between SALT2 and \pisco.


\subsection{\textit{B}-band peak magnitude comparison}
\label{subsec:b_band}

Before comparing \mb values, an offset of $\simeq0.27$\,mag needs to be applied as the version of SALT2 used in \citet{Scolnic18} incorporates this global offset. On close inspection, we also found that the MW extinction reddening values adopted by \citet{Scolnic18}, using the \citet{Schlafly11} dust maps, had some inconsistencies, principally in the low-$z$ sample where most SNe have smaller reddening values than expected. We use the same reddening values as \citet{Scolnic18} for comparison purposes.

\begin{figure*}
	\includegraphics[width=\columnwidth]{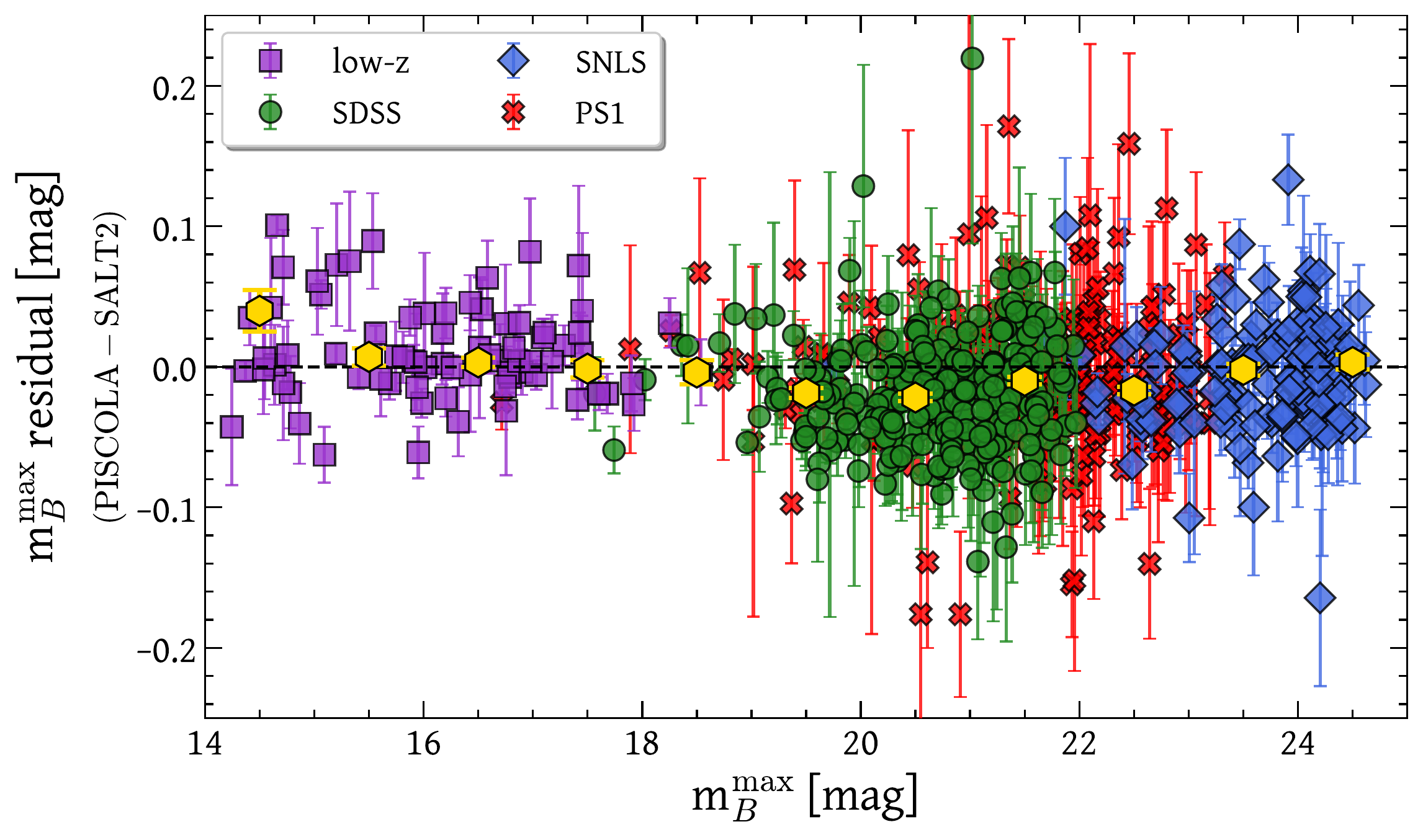}
	\includegraphics[width=\columnwidth]{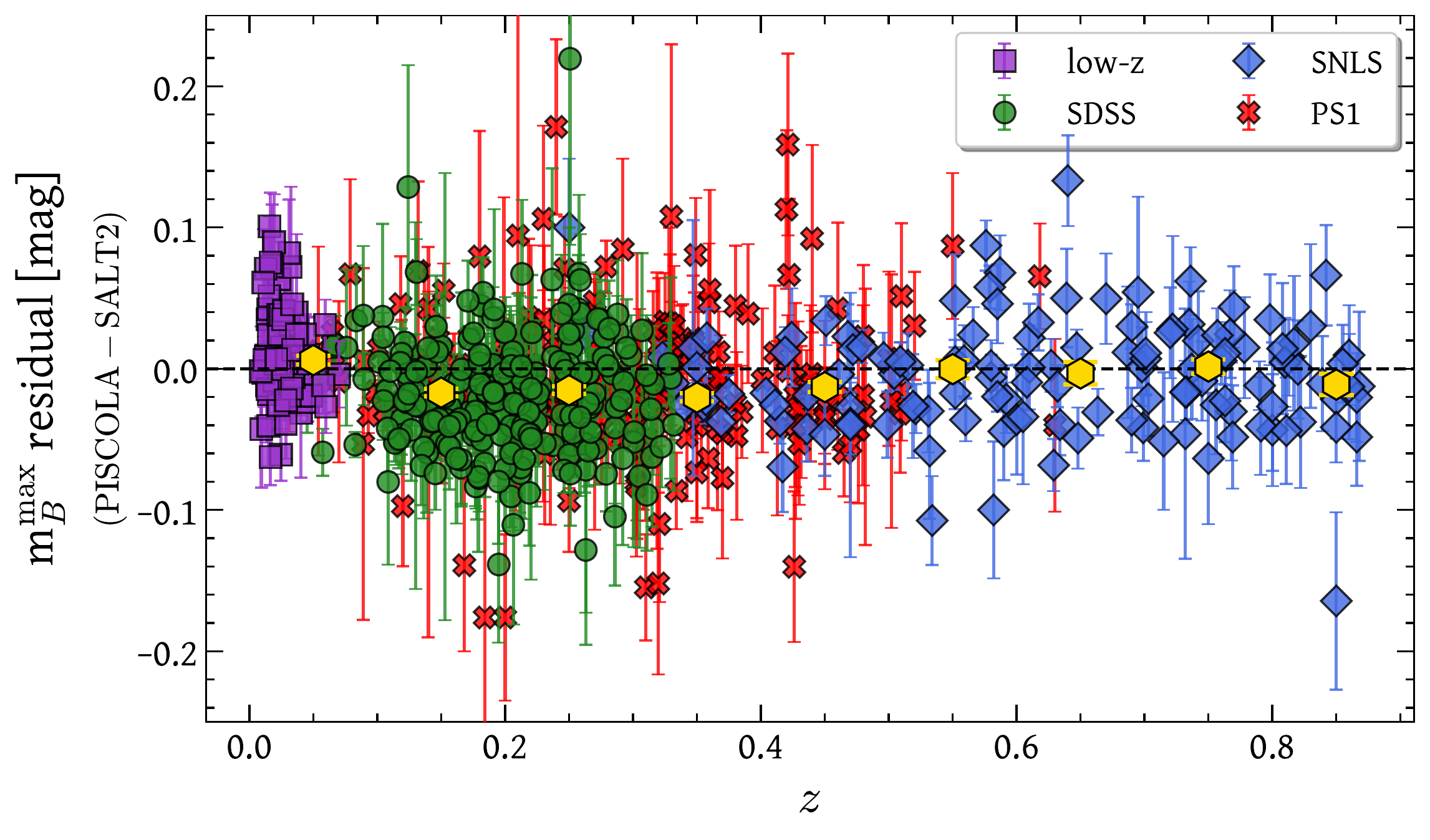}
    \caption{Comparison between \mb obtained from the SALT2 light-curve fitter and that obtained using \pisco as a function of \mb (\textbf{left} panel) and $z$ (\textbf{right} panel) for SNe Ia from the Pantheon sample. The error bars are 1$\sigma$ uncertainties from \pisco. The yellow hexagons represent the weighted mean in bins of 1\,mag (\textbf{left} panel) and 0.1 (\textbf{right} panel) with their respective uncertainties (1$\sigma$).}
    \label{fig:mb_comparison}
\end{figure*}

In Fig.~\ref{fig:mb_comparison}, we show the results of the comparison between \mb. The weighted average and weighted standard deviation of $\Delta\mathrm{m}^{\rm max}_{B} \equiv \Delta\mathrm{m}^{\rm max}_{B, \,\text{PISCOLA}}-\Delta\mathrm{m}^{\rm max}_{B, \,\text{SALT2}}$ are shown in Table~\ref{tab:pantheon_results}. There is a general formal disagreement ($>$3$\sigma$) for all the surveys except the low-$z$ sample.

\begin{center}
\begin{table}
\centering

\caption{Weighted mean, uncertainty on the weighted mean and weighted standard deviation of $\Delta$\mb and \colour$-~c$ for the Pantheon sub-sample used in this work (see Sec.~\ref{sec:salt2_comparison}).}

\begin{tabular}{cccc}
  &   &  error on the  &  weighted\\
   survey & weighted mean &  weighted mean &   standard deviation \\
     & (mmag) & (mmag) & (mmag) \\
\hline
\multicolumn{4}{c}{$\boldsymbol{\Delta\mathrm{m^{max}_{\textit{\textbf{B}}}}}$}\\
low-$z$       & 10        &    4            & 31                                         \\
SDSS          & -26       & 2                  & 31                                         \\
SNLS          & -8        & 2              & 25                                         \\
PS1           & -9       & 2                    & 34                                        \\
\hline
\multicolumn{4}{c}{$\boldsymbol{(B-V)_{\rm max} - c}$}\\
low-$z$       & 45            & 1               & 8                                        \\
SDSS          & 35         & 3                  & 38                                         \\
SNLS          & 9       & 5                    & 27                                         \\
PS1           & 4      & 4                    & 44                                           \\


\end{tabular}
\label{tab:pantheon_results}
\end{table}

\end{center}

The differences are typically small, but there are some important details. The apparent offset observed for the low-$z$ sample is mainly driven by one SN, SN\,2004ey (in the [$14, 15$]\,mag bin in the left panel of Fig.~\ref{fig:mb_comparison}), due to its large discrepancy with SALT2 (\mb residual of $\sim0.1$\,mag), but small uncertainties. If this object is removed, the weighted average of the \mb residual for the low-$z$ sample is reduced to $0.005$\,mag. Using SALT2 to fit this SN, we obtained relatively good fits to the light curves (see Fig.~\ref{fig:sn_fit_salt2}), but with a different value of \mb compared to that published by \citet{Scolnic18}, although in better agreement with the \pisco value. We do not know the exact source of this difference and do not have any reason from the \pisco fit, which results in better residuals than SALT2 (see Fig.~\ref{fig:sn_fit_pisco}), to discard this SN.

The discrepancy in \mb for the SDSS sample is the largest, expected due to its relatively larger uncertainties. The lower S/N may cause an apparent over-prediction of \pisco-measured \mb values for these SNe (see discussion in Section~\ref{subsec:testing_cadence}). However, there is also milder disagreement for the SNLS and PS1 surveys, perhaps due to unidentified issues with photometric calibration given the results of Section~\ref{sec:test}, or due to differences in the SED models between SALT2 and \pisco. Despite thoroughly checking the analysis and our tests, we are unable to identify the source of this discrepancy. We note that the differences could be due to unidentified issues in \pisco, SALT2 or both.

\begin{center}

\begin{table*}
\caption{Number of supernovae discarded at different stages in the \pisco fitting.}
\centering
\begin{tabular}{ccccccc}
Discarding reason & low-$z$ & SDSS & SNLS & PS1   & Total & Cumulative number  \\ 
&&&&&&discarded\\
\hline


Poor wavelength coverage$^{a}$             & 1     & 3     & 33    & 9     & 46    & 46    \\
Poor time coverage$^{b}$             & 58    & 21    & 7     & 11     & 97    & 143    \\
Inaccurate results$^{c}$             & 22    & 18    & 1     & 13    & 54    & 197   \\
Not satisfying extra constraints$^{d}$             & 15    & 39    & 57    & 32    & 143   & 340    \\
Poor light-curve fits$^{e}$             & 2     & 24    & 8     & 28    & 62     & 402  \\
\hline
Initial sample  & 172   & 335   & 236   & 279   & 1022  &   \\ 
\hline
Total discarded SNe   & 98    & 115    & 106    & 93    & 402   &    \\
\hline
Remaining SNe   & 74     & 230     & 130     & 186     & 620 &
\end{tabular}

\textbf{Notes.} The reasons for discarding SNe are: ($a$) poor wavelength coverage that does not allow the calculation of the rest-frame $B$-band light curve, ($b$) poor time coverage that does not allow an initial estimation of \tmax, ($c$) unable to estimate an accurate $B$-band light curve after light-curve fit and correction, ($d$) unable to satisfy extra constraints on peak coverage (see the penultimate paragraph of Sec.~\ref{sec:test}), and ($e$) poor \pisco fits checked by visual inspection of SNe with large discrepancy in \mb compared to SALT2 values.
\label{tab:sample_cuts}
\end{table*}
\end{center}

\subsection{Colour comparison}
\label{subsec:colour}

\begin{figure}
	\includegraphics[width=\columnwidth]{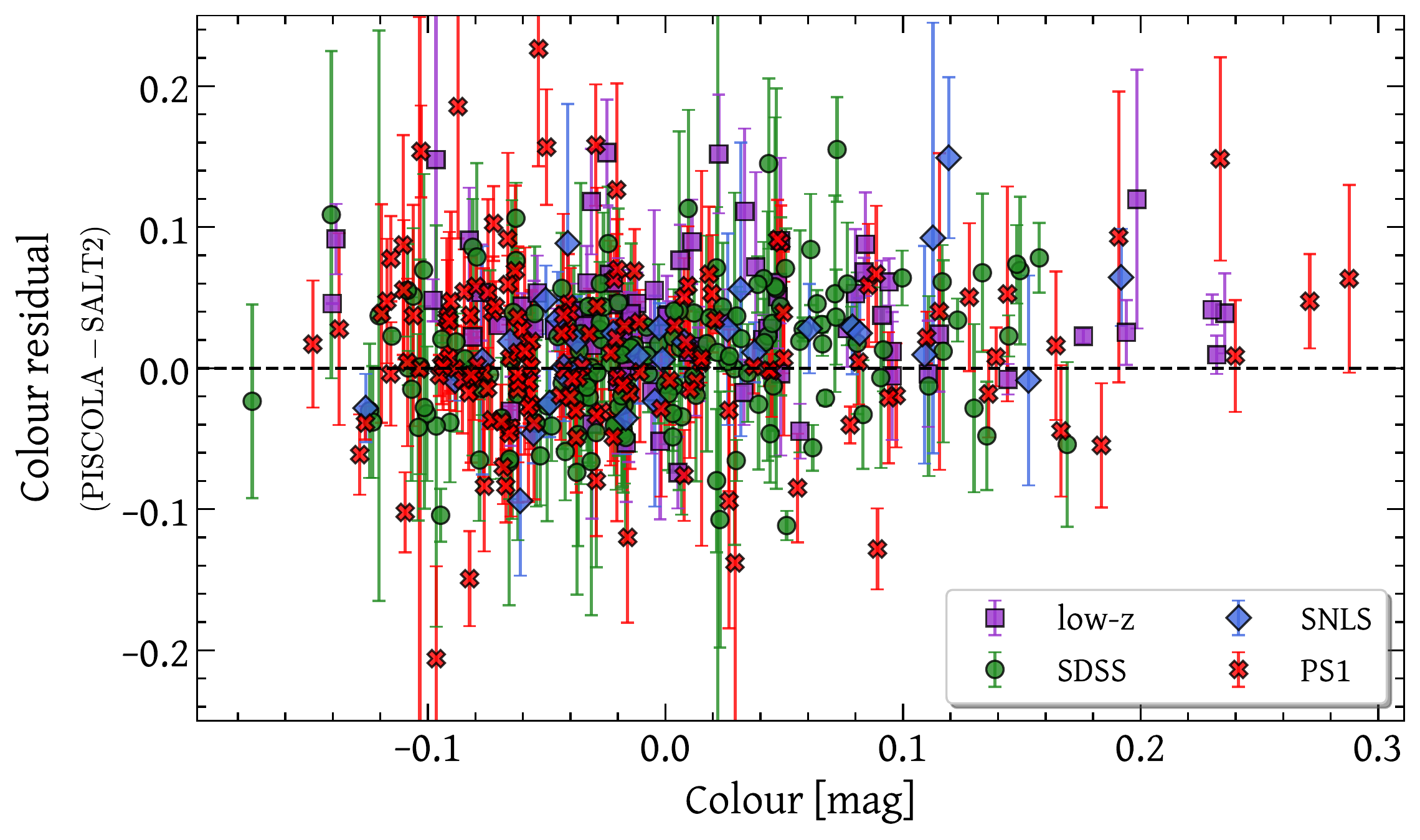}
    \caption{Colour comparison between \pisco \colour and SALT2 $c$. The error bars are 1$\sigma$ uncertainties from \pisco. Note that $c$ is a more indirect measurement of \colour than the \pisco estimator.}
    \label{fig:colour_comparison}
\end{figure}

We compare the colour parameters in Fig.~\ref{fig:colour_comparison}. Given the data-driven nature of \pisco, not every SN with an estimation of \mb has sufficient wavelength coverage to also estimate its \colour, particularly at high redshift. We also note that the colour parameters are fundamentally different \citep[see, e.g.,][]{Kessler13}, and thus a detailed comparison is difficult. For exampled, SALT2 estimates \colour through the $c$ parameter using information across the 3000--7000\,\angstrom range, while \pisco makes a direct measurement of $B-V$.

The weighted average and weighted standard deviation for the colour parameter residuals are shown in Table~\ref{tab:pantheon_results}. As expected, some differences are seen between the colour parameters, particularly in the low-$z$ and SDSS samples, where \pisco measures redder SNe than SALT2. A global offset may be expected, but the discrepancies are different for each survey, possibly implying issues with calibration as for \mb, or with the SED model.

In summary for this section, we find some differences between light-curve parameters for SNe Ia measured using \pisco and SALT2, some of which were anticipated. For the most straight forward comparison, i.e. \mb, the differences between \pisco and SALT2 were small. A detailed comparison for colour is more difficult due to slightly varying definitions. Taking the results from this section and from Section~\ref{sec:test}, we conclude that the performance of \pisco is satisfactory and validates our code.

\section{SN Ia Light Curve Analysis}
\label{sec:lc_analysis}

Having introduced \pisco as a general light-curve fitter tool for SNe, we now demonstrate a simple use case applied to the standardisaton of SNe Ia for distance estimation. Many current and previous cosmological analyses \citep[e.g.,][]{Astier06,Guy10, Betoule14, Scolnic18} have used SALT/SALT2 light-curve parametrizations to standardise SNe Ia using the Tripp-like formula \citep{Tripp98}:

\begin{equation}
    \mu=m_{B}-M+\alpha \times x_{1}-\beta \times c.
    \label{eq:tripp_original}
\end{equation}

where $\mu$ is the SN Ia distance modulus, $m_{B}$, $x_1$ and $c$ are the SALT2 light-curve parameters, and $\alpha$, $\beta$ and $M$ are nuisance parameters. $M$ represents the average $B$-band peak absolute magnitude of SNe Ia. 

Current cosmological analyses using equation~\ref{eq:tripp_original} incorporate an additional intrinsic dispersion term ($\sigma_{\rm int}$), which encapsulates additional SN Ia variability that cannot be explained by their standardisation \citep[e.g.,][]{Perlmutter97, Tonry03, Riess04, Guy10, Betoule14, Scolnic18}. In the following subsections we analyse SN Ia light curves from the Pantheon sample using Non-negative Matrix Factorization (NMF), an unsupervised machine-learning method, in search of an alternative parametrizations to study the standardisation of SNe Ia.

\subsection{Light-Curve Decomposition with Non-negative Matrix Factorization}
\label{subsec:lc_decomp}

Previous works have used PCA (or variations of it) on SNe Ia to analyse their light curves and spectra to better understand their use as cosmological probes \citep[e.g.,][]{Cormier11, Kim13, Sasdelli16, He18, Saunders18}. Here we use a different approach and analyse our rest-frame $B$-band SN Ia light curves with NMF, a data-driven linear decomposition algorithm, comparable to PCA. NMF factorises a non-negative matrix, \textbf{\textit{X}}, with dimensions $m \times n$, in a set of two non-negative matrices, \textbf{\textit{W}} and \textbf{\textit{H}}, with dimensions $m \times k$ and $k \times n$ (where $k < min(n, m)$), respectively, such that \textbf{\textit{X}} $=$ \textbf{\textit{W}} $\times$ \textbf{\textit{H}}. In other words, NMF factorises a matrix into several components (eigen-vectors) and coefficients (eigen-values). Both methods (NMF and PCA) are commonly used for dimensionality reduction and feature extraction; however, NMF is better suited for some applications in astronomy as most astrophysical signals are non-negative, thus extracting components that are easier to interpret physically. Additionally, NMF components are not necessarily orthogonal, as is the case for PCA components, which allows us to find correlations between coefficients.

\begin{figure}
	\includegraphics[width=\columnwidth]{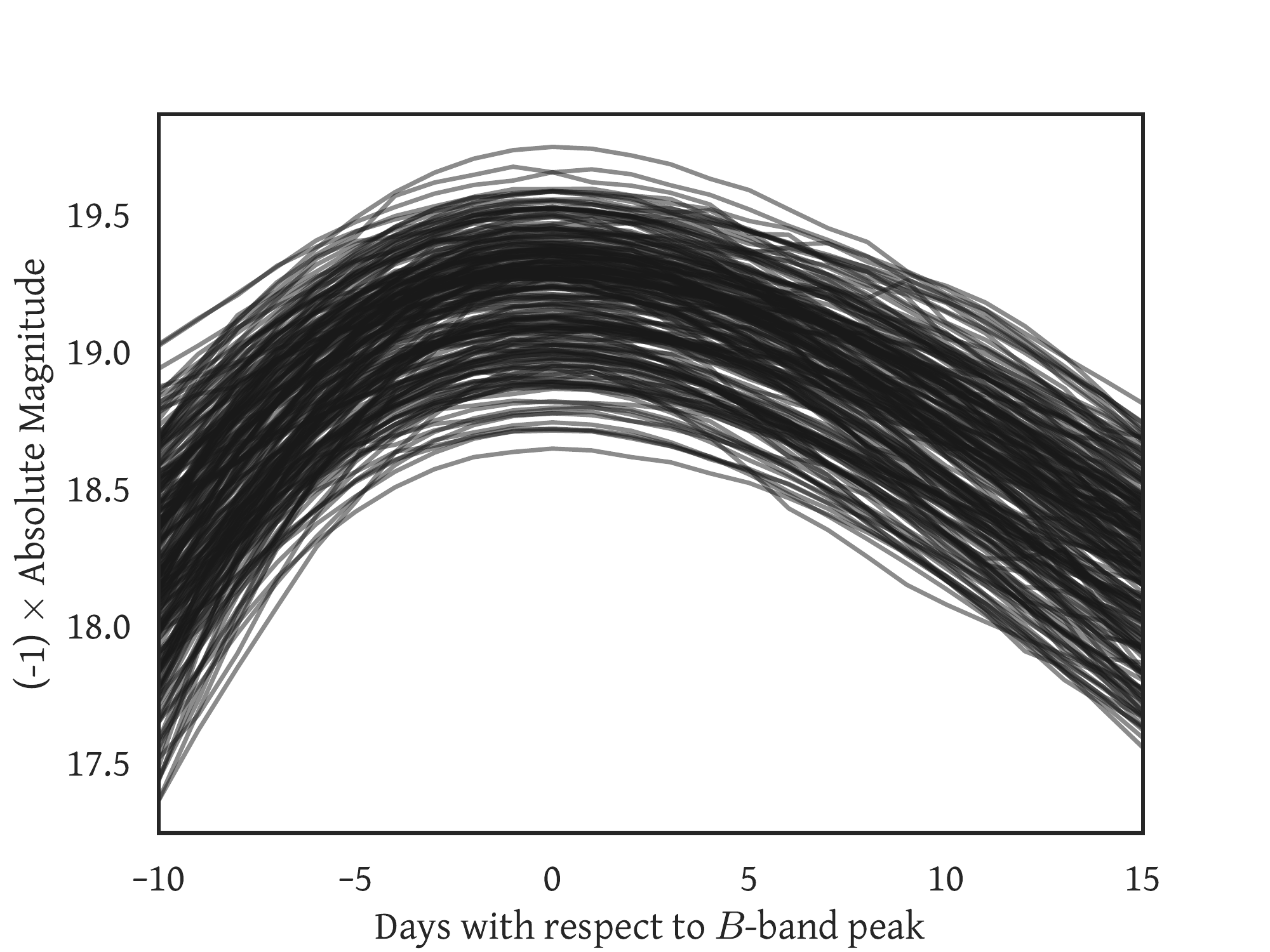}
    \caption{$B$-band light curves of 214 \pisco-fitted SNe Ia used in the decomposition analysis. Values are in absolute-magnitude multiplied by $-1$.}
    \label{fig:blcs}
\end{figure}

The light-curve decomposition depends on the phase coverage being used. We present our analysis for 214 SNe which have rest-frame data over the phase range $-$10 to $+$15\,d. In this proof-of-concept investigation, we also discarded 50 SNe which did not have straight-forwardly rising and then declining light curves, flagged by visual inspection of the fits. There are many reasons why such fits may result, including astrophysical reasons such as the presence of secondary peaks in SN Ia light curves, or experimental reasons such as lower S/N data. However, this decomposition application is deliberately designed to be simple in scope; future work will examine the more complicated morphology of SN Ia light curves. The sample selection for this analysis is summarised in Table~\ref{tab:analysis_cuts}. In Section.~\ref{subsec:further_exploration}, we explain in more detail the phase range chosen, and explore other phase ranges. The rest-frame $B$-band light curves for this sample are shown in Fig.~\ref{fig:blcs}.

We decomposed these light curves in absolute-magnitude space, multiplied by $-1$ to obtain positive values. To calculate the absolute magnitudes we use distance modulii calculated with our assumed cosmology and the redshifts from \citet[][specifically $z$HD]{Scolnic18}. As the NMF algorithm used does not incorporate uncertainties, we used a Monte Carlo approach, generating sets of light curves for each SN from the uncertainties estimated by \pisco, and applying NMF decomposition to each of these sets. This generates a distribution of coefficients from which we used the mean value and the standard deviation to propagate uncertainties.

\begin{figure}
	\includegraphics[width=\columnwidth]{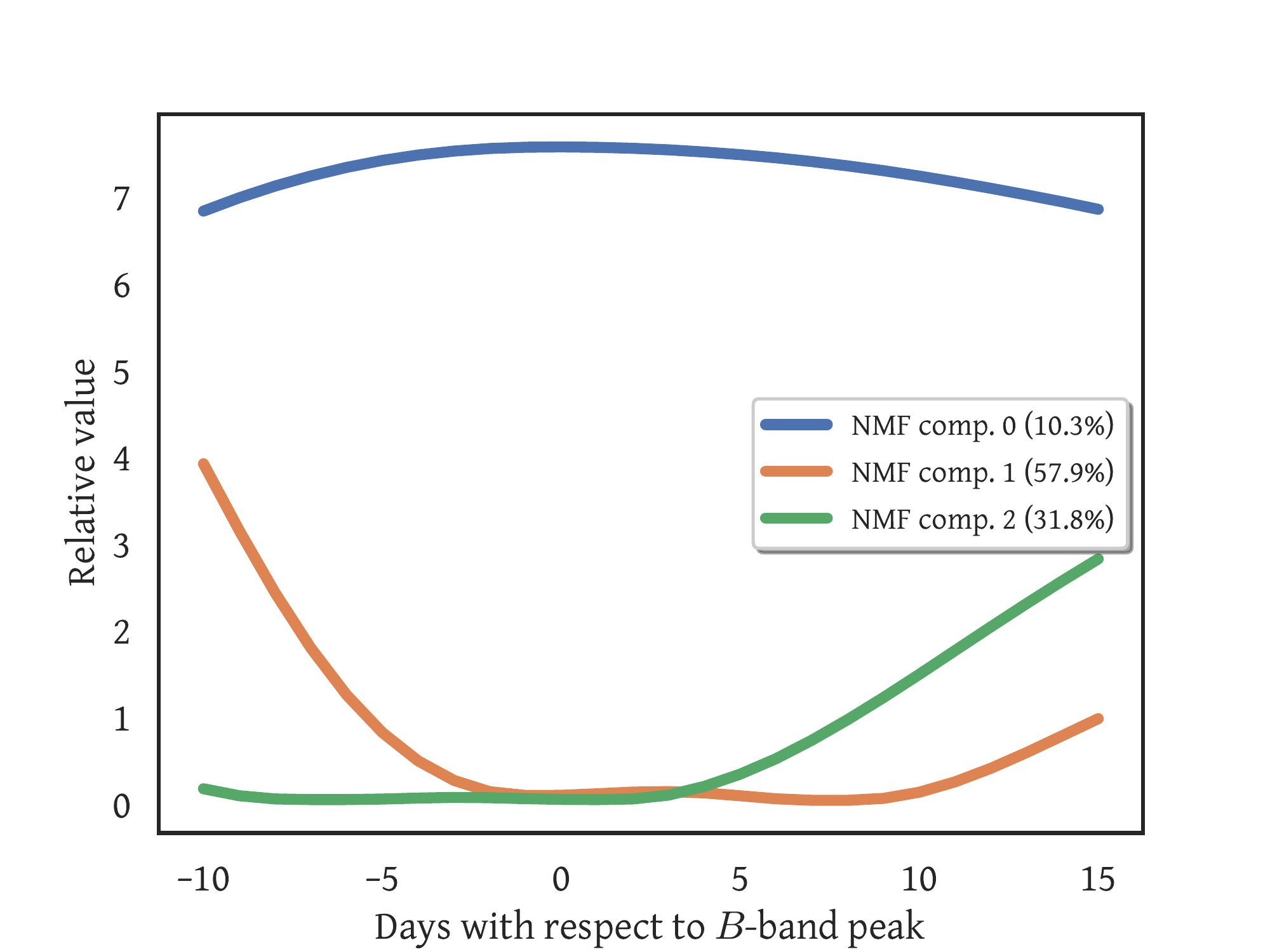}
    \caption{Three NMF components obtained from the decomposition of the $B$-band light curves shown in Fig.~\ref{fig:blcs}. In parenthesis are the percentages of the explained variance for each component with respect to the total variance of all components. Note that the absolute values of the components are not important for this analysis.}
    \label{fig:nmf_components}
\end{figure}

The following analysis is described for a decomposition with three components, although we consider further components in Section~\ref{subsec:further_exploration}. In Fig.~\ref{fig:nmf_components}, we show the components obtained with NMF, with their respective explained variance (in percentage) with respect to the total variance of all components. Each eigen-vector contains specific information about the $B$-band of an \lq average\rq\ SN Ia. Component $0$ contributes to the general scale of the light curve, thus correlating with the $B$-band peak absolute magnitude, \Mb. Component $1$ contributes to the rise of the light curve, while component $2$ mainly contributes to the decline of the light curve (and therefore we expect it to correlate with \dm). Component $1$ also contributes to the decline, but to a lesser degree than component $2$. We note that from this data-driven decomposition, we naturally retrieve components related to the rise and decline of the $B$-band light curve, in agreement with the findings of \citet{Hayden10} and \citet{Ganeshalingam11}.

We are principally interested in the relative values of the components rather than the absolute values, which have no direct physical interpretation. We label the coefficients $p_0$, $p_1$ and $p_2$, associated to components $0$, $1$ and $2$, respectively. 

\begin{figure}
	\includegraphics[width=\columnwidth]{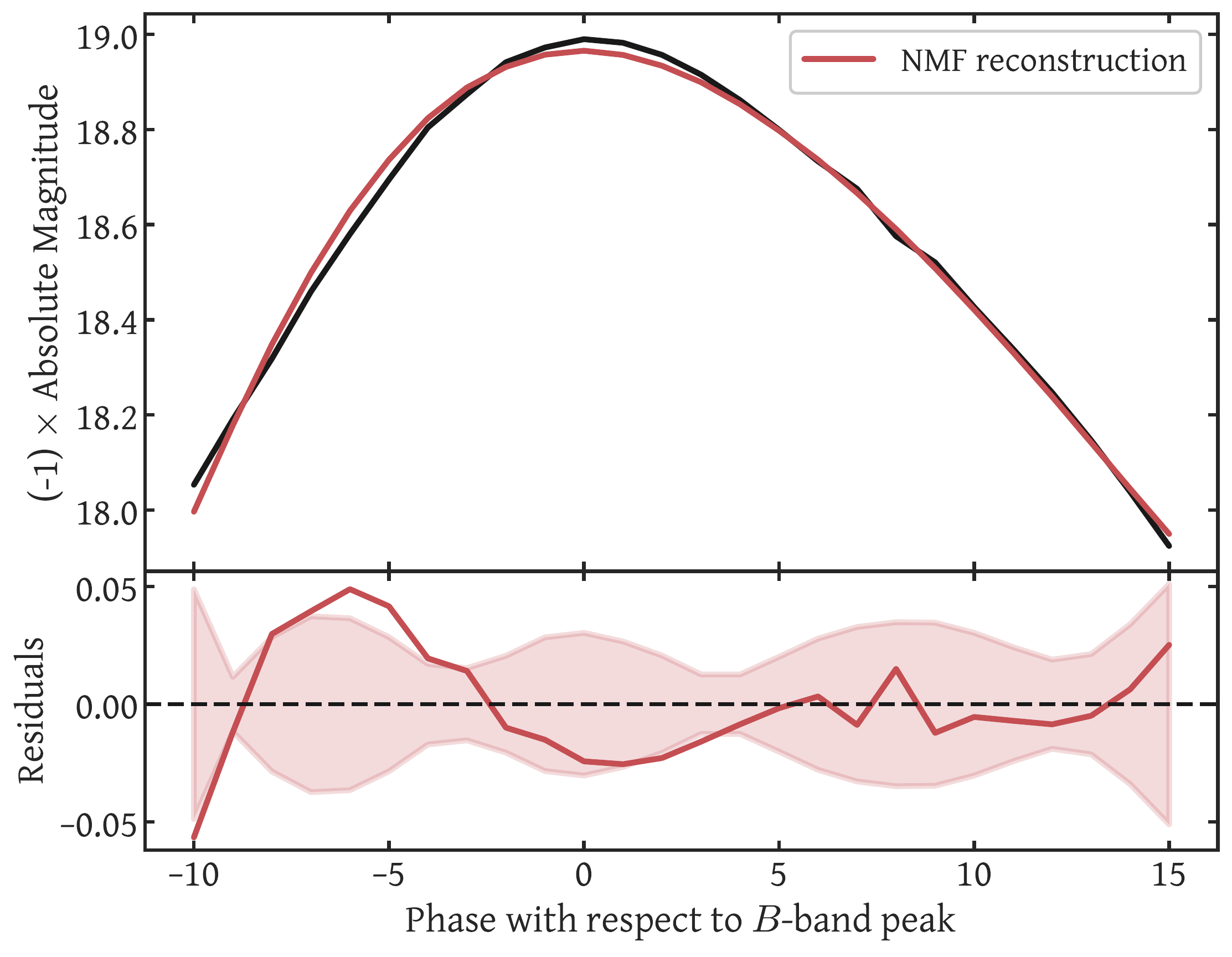}
    \caption{\textbf{Top:} example of $B$-band light curve from one of the SNe Ia (black) together with a reconstructed light-curve using the NMF components and coefficients (red). Values are in absolute-magnitude multiplied by $-$1. \textbf{Bottom:} residuals from the light-curves shown in the top panel. The shaded area shows the standard deviation of the residuals for the entire sample in Fig.~\ref{fig:blcs}.}
    \label{fig:nmf_reconstruction}
\end{figure}

In Fig.~\ref{fig:nmf_reconstruction}, we show an example of a $B$-band light curve from one of the SNe together with a reconstructed light curve using the NMF components and coefficients, and their residuals. The residuals show that the reconstructed light curve has small differences ($\lesssim$0.05\,mag) to the original light-curve during the rise, but better agreement around peak ($\sim$0.02\,mag residuals) and at later times ($\sim$0.00\,mag residuals). The reconstructed light-curves and the original light-curves for the sample of 214 SNe are generally in excellent agreement, with mean residuals of $\sim$0.0\,mag and a standard deviation of $\sim$0.03\,mag for all phases. This demonstrates that the NMF decomposition with three components is able to capture the variation in the light curves of SNe Ia.

\begin{figure}
	\includegraphics[width=\columnwidth]{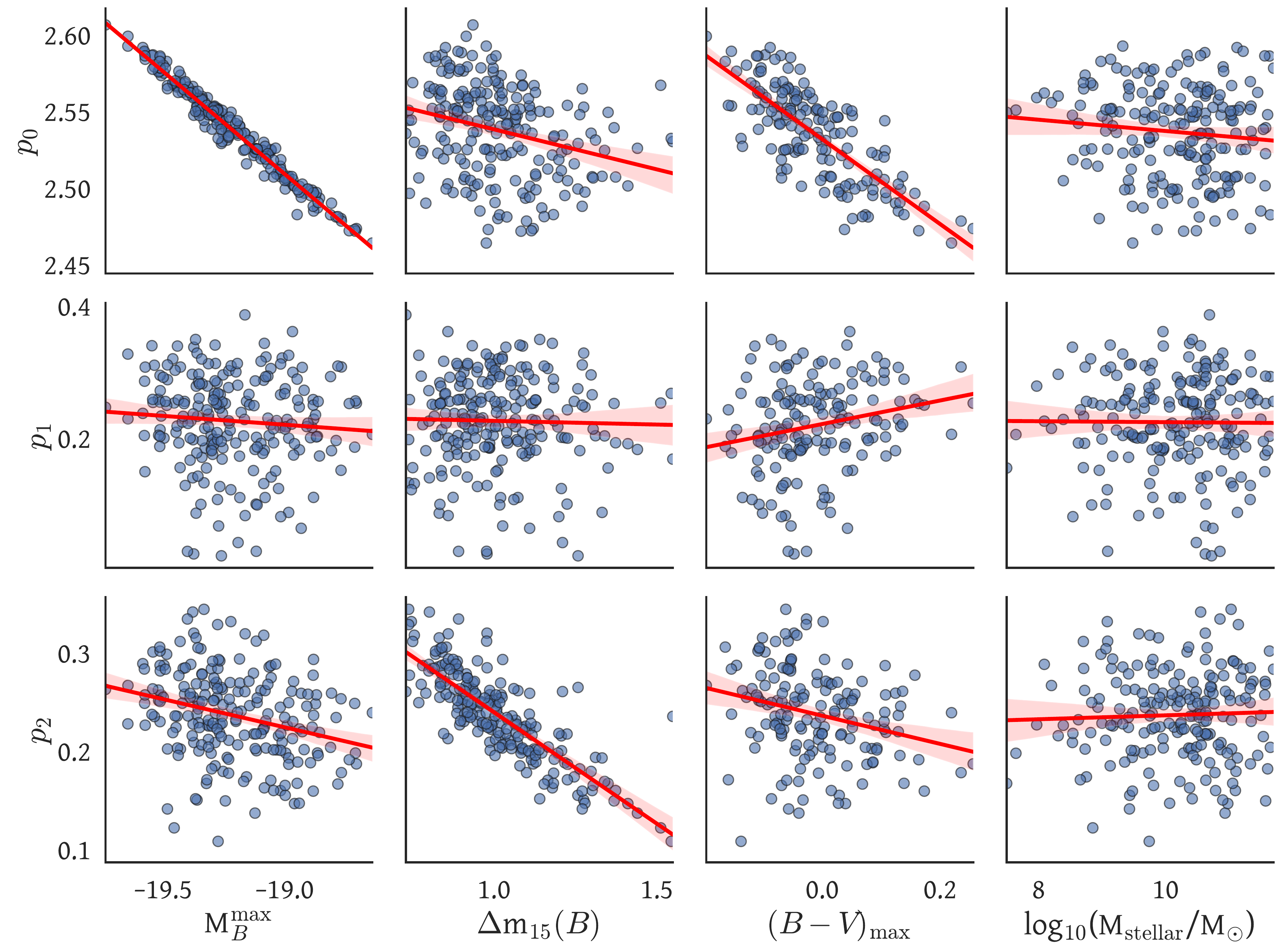}
    \caption{NMF coefficients vs \pisco light-curve parameters and host galaxy stellar mass for the SNe in Fig.~\ref{fig:blcs}. From the 214 SNe, only 156 have \colour values. Uncertainties are not shown for visualisation. Linear regressions are represented by red lines.}
    \label{fig:coef_vs_lcparams}
\end{figure}

The coefficients tell us about the contribution (or weight) of each of the components on the light curve of a SN Ia. By comparing these with different light-curve parameters, we can better understand their physical interpretation. This is shown in Fig.~\ref{fig:coef_vs_lcparams}. As  expected, there is a clear correlation between $p_0$ and \Mb, and thus \colour \citep[e.g.,][]{Tripp98}. Coefficient $p_1$, which contributes to the rise of the light curve, has a small correlation with \colour, but not with \Mb. On the other hand, coefficient $p_2$ clearly correlates with \dm, as expected, and shows minor correlations with \Mb and \colour. None of the components correlates with host-galaxy stellar mass. The decomposition is somewhat analogous to the SALT2 model, which contains $x_0$ and $x_1$ terms, although here we have two stretch components. In our case, the average ($B$-band light-curve) model has stretch parameters $p_1 \sim p_2 \sim 0.2$, while in the case of SALT2 it has $x_1 = 0$.

We note that NMF, unlike PCA, does not produce orthogonal components given the constraints of non-negative values, i.e., correlations between the components can be expected. From the coefficients, we find that $p_1$ and $p_2$ are anti-correlated (correlation of $\sim -$0.5).


\subsection{Distance estimation}
\label{subsec:mod_tripp}

We follow an analogous approach to equation~\ref{eq:tripp_original} using the \pisco components:

\begin{equation}
    \mu=m_{B}-M+\eta_1 \times \hat{p}_1+\eta_2 \times \hat{p}_2-\beta \times (B-V)_{\rm max}
\end{equation}

where $m_B$ and \colour are the \pisco-measured light-curve parameters (i.e., measured from the fits), $\hat{p} = p - \langle p\rangle$, and $\langle p\rangle$ is the average NMF coefficient value of the sample.

\subsubsection{Hubble Diagram}
\label{subsubsec:hd}

To build a Hubble diagram, we adopt our standard cosmology (see Section~\ref{sec:intro}) and use a Markov Chain Monte Carlo (MCMC) to find the optimal values for the nuisance parameters $M$, $\eta_1$ , $\eta_2$ and $\beta$, adopting uniform priors without bounds (except for $M$, where we require $M < 0$). Note that we used only SNe with \colour values (156 out of 214) for the Hubble diagram. See Table~\ref{tab:analysis_cuts} for more details regarding the cuts applied to the SNe sample and the end of Section~\ref{subsec:further_exploration} for a summary. The need of colour limits the range of redshift, as we require rest-frame $V$-band coverage to estimate \colour. Future work will focus on alternative colour parameters. For instance, the slope of the mangling function contains information about \colour, which is approximately contained in a narrow range ($\sim$ 4500--5500\,\angstrom), allowing us to incorporate higher-redshift SNe, although it provides a more limited standardisation compared to \colour.

\begin{center}
\begin{table*}
\caption{Number of SNe Ia discarded at different stages of the analysis (Section~\ref{sec:lc_analysis}).}
\centering
\begin{tabular}{ccccccc}
Discarding reason & Low-$z$ & SDSS & SNLS & PS1   & Total & Cumulative number  \\ 
&&&&&&discarded\\
\hline
Incomplete phase coverage           & 59    & 129   & 71    & 97        & 356   & 356   \\
Unphysical light curve     & 1     & 19    & 8     & 22        & 50    & 406   \\
No \colour estimation               & 0     & 12    & 40    & 6         & 58    & 464   \\
\hline
Initial sample                      & 74    & 230   & 130   & 186       & 620   &   \\ 
\hline
Total discarded                     & 60    & 160   & 119   & 125       & 464   &    \\
\hline
Remaining SNe                       & 14    & 70    & 11    & 61        & 156   &
\end{tabular}

\label{tab:analysis_cuts}
\end{table*}
\end{center}

\begin{figure}
	\includegraphics[width=\columnwidth]{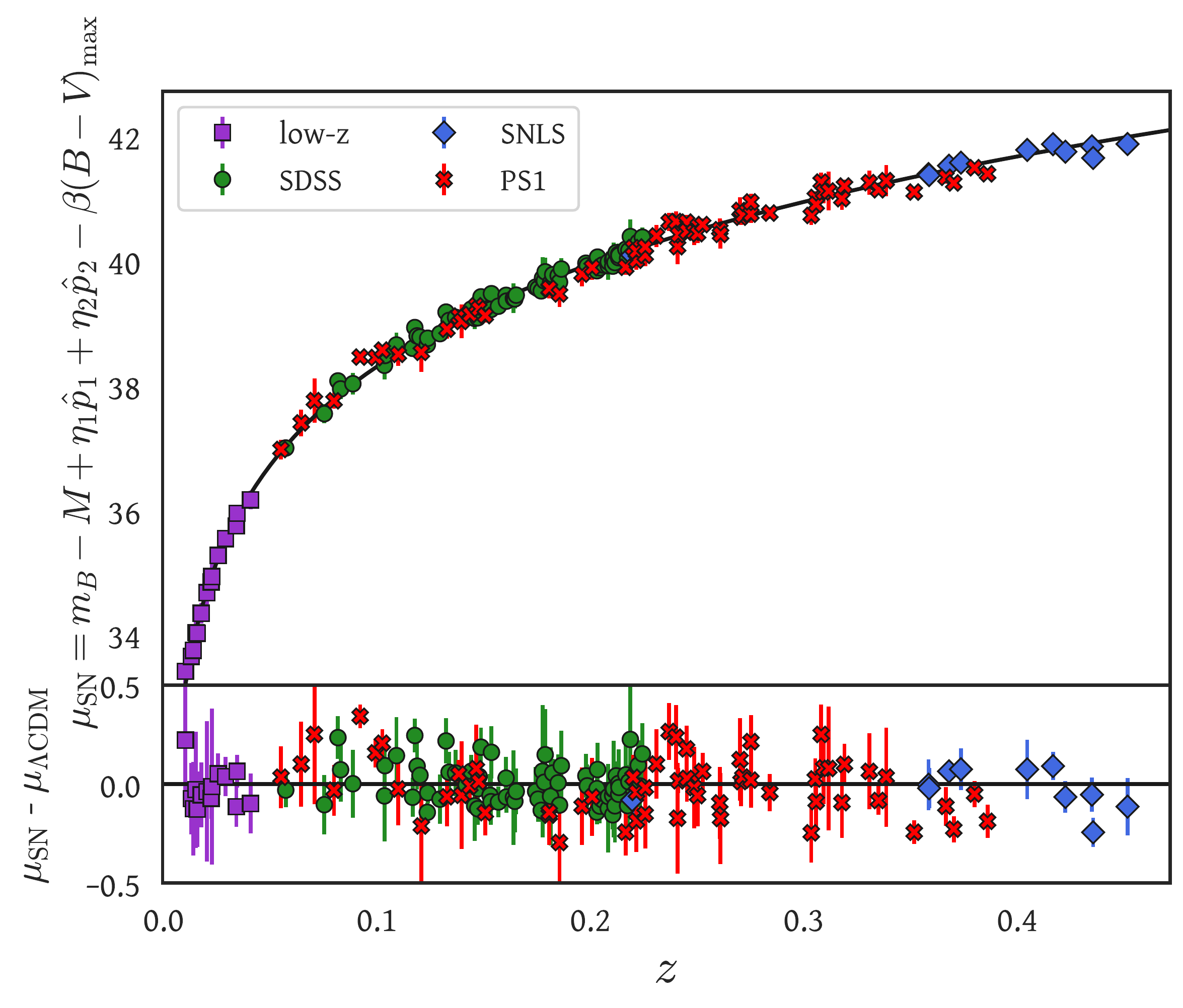}
    \caption{Hubble diagram (\textbf{top} panel) and residual (\textbf{bottom} panel) for the SNe Ia in the sample, using the \pisco standardisation introduced in this work.}
    \label{fig:hubble_diagram}
\end{figure}

\begin{figure}
	\includegraphics[width=\columnwidth]{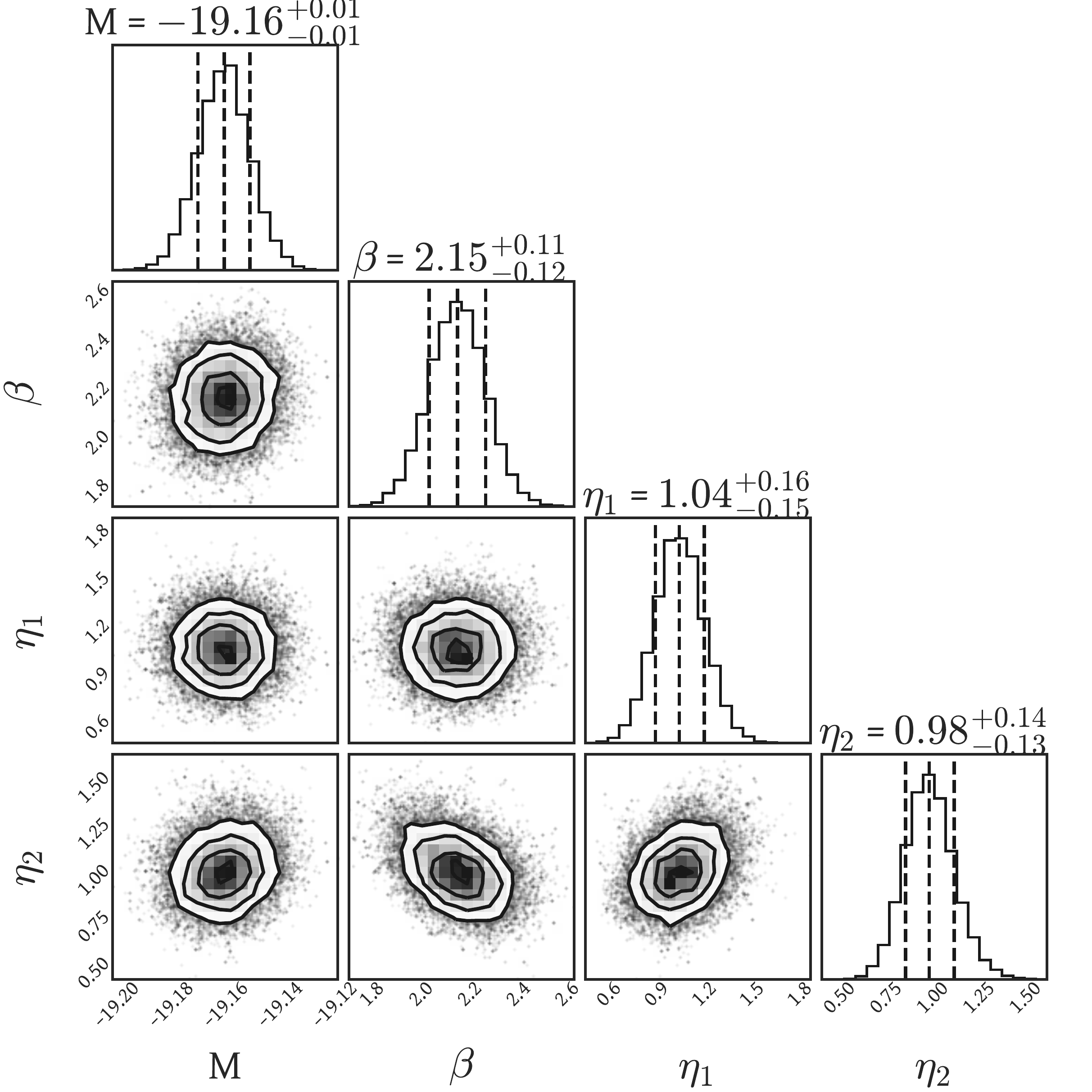}
    \caption{MCMC results of the nuisance parameters used for the standardisation of the SNe Ia in the Hubble diagram (Fig.~\ref{fig:hubble_diagram}).}
    \label{fig:mcmc}
\end{figure}

We included uncertainties and covariances for our light-curve parameters, uncertainties in redshift ($\sigma_z$), uncertainties due to peculiar velocities ($\sigma_{\rm pec}$; adopting 300 km~s$^{-1}$) and uncertainties from stochastic gravitational lensing ($\sigma_{\rm lens}$). We also included an intrinsic scatter term ($\sigma_{\rm int}$) such that $\chi^2_{\rm red} = 1$. Bias corrections are not included as a separate analysis would be required. However, we do not expect it to be significant due to the need for a \colour measurement in our analysis, which translates into an implicit redshift cut; see the redshift ranges for the different surveys in Fig.~\ref{fig:hubble_diagram}. The resulting Hubble diagram is shown in Fig.~\ref{fig:hubble_diagram}, while the MCMC results of the parameters are in Fig.~\ref{fig:mcmc}. 

\subsubsection{Hubble Residuals: PISCOLA vs SALT2}
\label{subsubsec:hr}

\begin{figure}
	\includegraphics[width=\columnwidth]{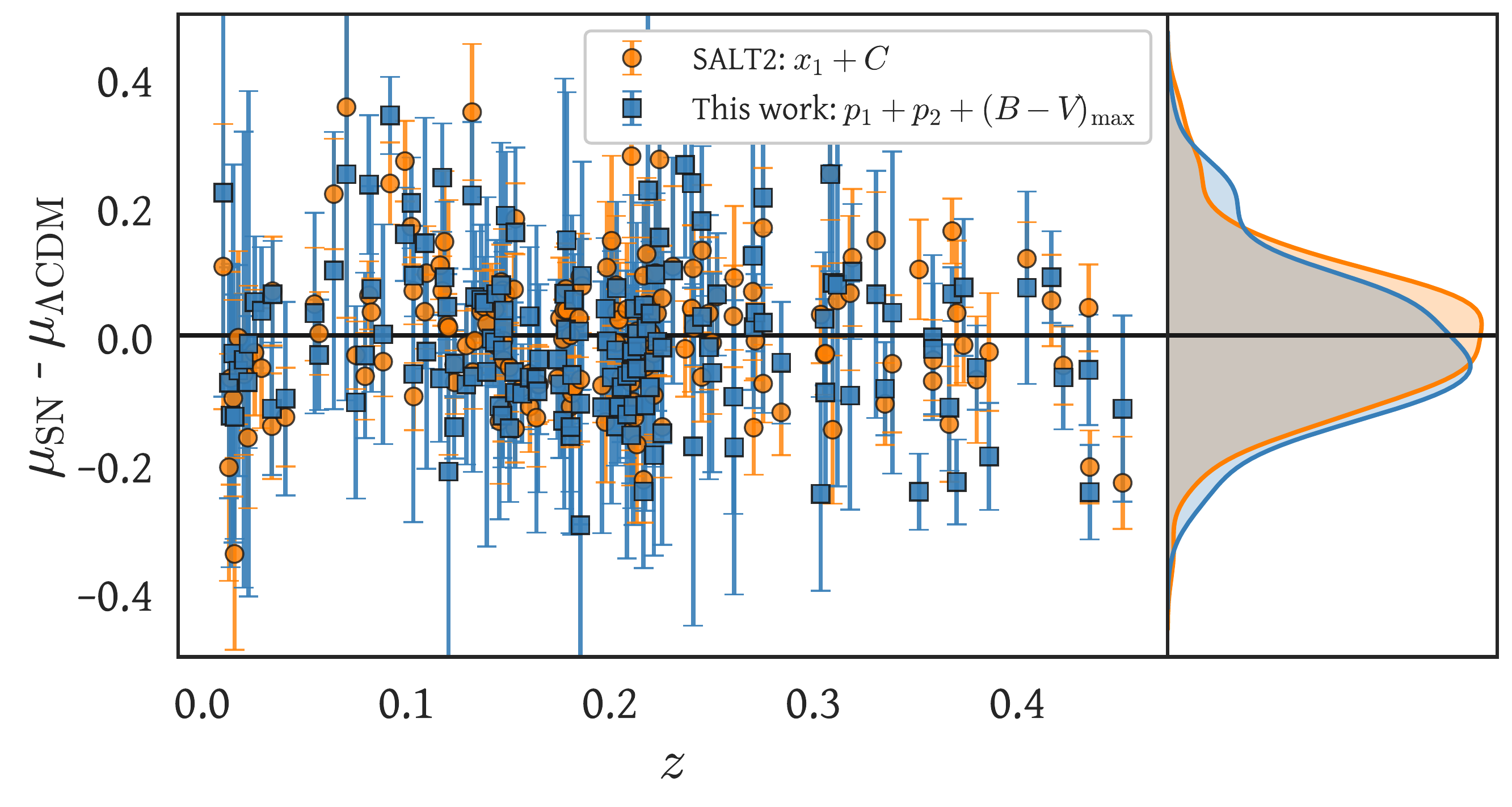}
    \caption{Hubble Residual comparison between the parametrization derived in this work (r.m.s. $=$ 0.118\,mag) and the standard parametrization from SALT2 (r.m.s. $=$ 0.111\,mag).}
    \label{fig:hubble_residuals}
\end{figure}

We now compare this simple \pisco parametrization with SALT2 using the same sample of 156 SNe. In Fig.~\ref{fig:hubble_residuals}, we show the comparison of the Hubble residuals measured using parameters derived from the \pisco light curves and from SALT2. \pisco obtains a similar r.m.s. value of $0.118$\,mag compared to SALT2 ($0.111$\,mag), indicating it as a competitive method. The performance difference in r.m.s. of 0.007\,mag is small. We note, however, that, although \pisco does not outperform SALT2, using it for cosmological analyses is a limited demonstration of its  potential as a general purpose light-curve fitter.

The nuisance parameters obtained by using \pisco and SALT2 are summarised in Table \ref{tab:cosmo_params}. For SALT2, we found a similar value for $\alpha$ and a slightly lower value for $\beta$ than those reported in \citet{Scolnic18}, possibly due to the subsample used in this work. The value for $M$ is different between both approaches, but as a normalisation factor this has no effect on the analysis. The value of $\beta$ is smaller with \pisco, indicating the colour correction is not as large as with SALT2, possibly implying that some of the colour contribution is accounted for in $p_1$ and/or $p_2$ (note also the slight correlation between $\beta$ and $\eta_2$ in Fig.~\ref{fig:mcmc}). The stretch-like parameters, $p_1$ and $p_2$, distribute as quasi-Gaussians over $\sim0.0-0.4$ with mean values of $\sim0.2$, so the correction from $\eta_1p_1 + \eta_2p_2$ is larger than $\alpha x_1$. This is consistent with the possibility that some of the contribution from colour is absorbed by the $p_1$ and/or $p_2$ parameters. Finally, we obtain a lower $\sigma_{\rm int}$ with \pisco than with SALT2, but this is also a reflection of our larger uncertainties, where the main contributor to the uncertainty budget is \colour, an effect of the GP light-curve fits.

\begin{center}
\renewcommand{\arraystretch}{1.3}
\begin{table}
\caption{Nuisance parameters from the cosmological analysis for SALT2 and \pisco.}
\centering
\begin{tabular}{c|cc}
Parameter              & SALT2  & \pisco \\
\hline
$M$         & -19.36$^{+0.01}_{-0.01}$  & -19.16$^{+0.01}_{-0.01}$    \\
$\beta$     & 2.85$^{+0.08}_{-0.08}$    & 2.15$^{+0.11}_{-0.12}$      \\
$\alpha$    & 0.14$^{+0.01}_{-0.01}$    & -         \\
$\eta_1$    & -                         & 1.04$^{+0.16}_{-0.15}$      \\
$\eta_2$    & -                         & 0.98$^{+0.14}_{-0.13}$     \\
$\sigma_{\rm int}$ & 0.068              & 0.047  \\

\end{tabular}
\item \textbf{Notes.} The uncertainties in $\sigma_{\rm int}$ are negligible ($<$0.001) in both cases.

\label{tab:cosmo_params}
\end{table}
\end{center}

\begin{figure}
	\includegraphics[width=\columnwidth]{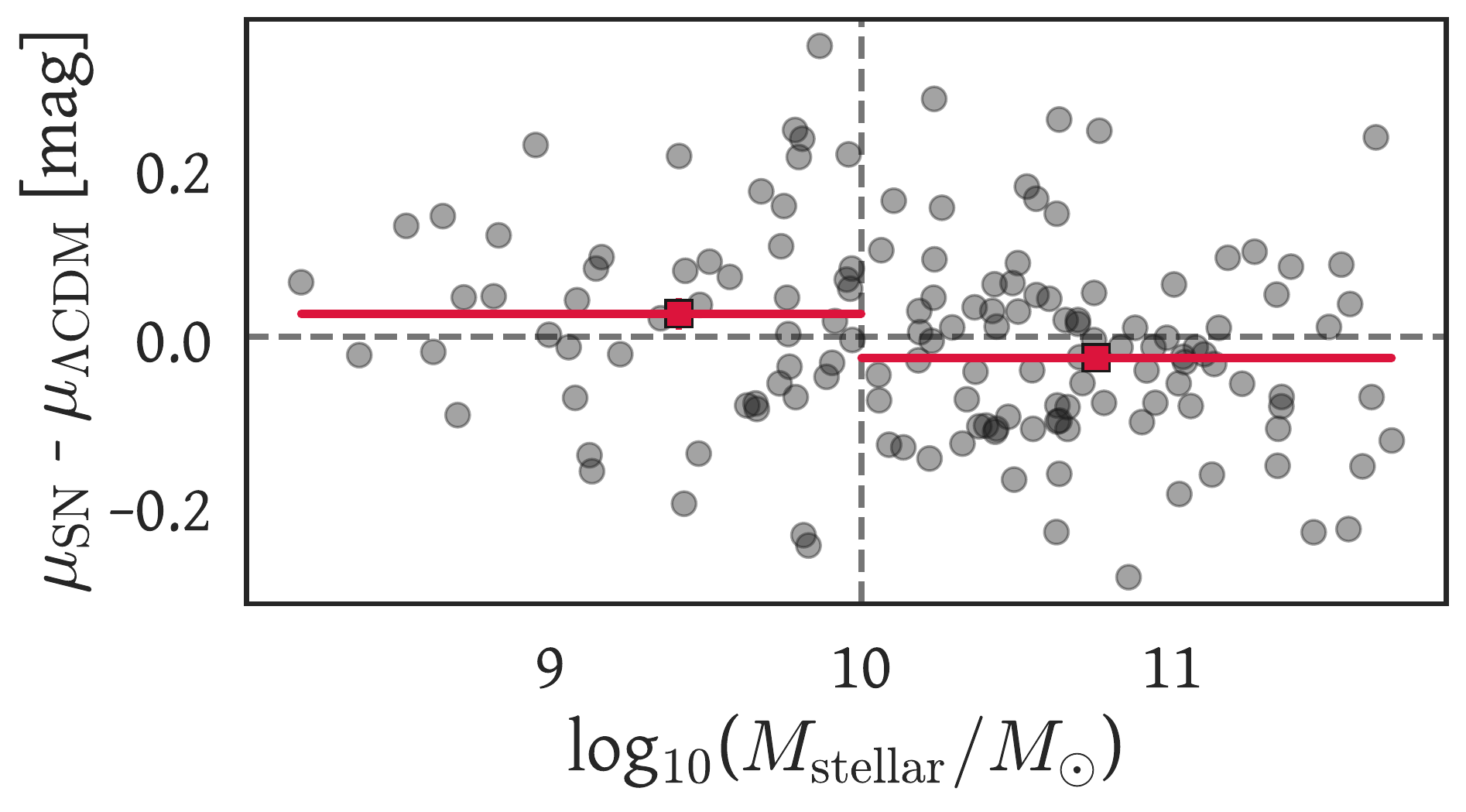}
    \caption{\pisco Hubble residuals as a function of host galaxy stellar mass. Uncertainties are not shown for visualisation. A host galaxy mass step of $-0.052 \pm 0.022$ mag ($2.4 \sigma$) was obtained. The vertical dashed line marks the location of the step at $M_{\rm stellar} = 10^{10} M_{\odot}$.}
    \label{fig:mass_step}
\end{figure}

We also examined the dependence of Hubble residual on host galaxy stellar mass, the so-called \lq mass step\rq\ \citep[e.g.,][]{Kelly10, Lampeitl10, Sullivan10}, by using the stellar mass values from \citet{Scolnic18}. We found a mass-step value of $-0.052 \pm 0.022$\,mag ($2.4 \sigma$; see Fig.~\ref{fig:mass_step}) and $-0.071 \pm 0.016$\,mag ($4.4 \sigma$) for \pisco and SALT2, respectively, for a step at $M_{\rm stellar} = 10^{10} M_{\odot}$. The \pisco value is therefore consistent with the value obtained with SALT2 and with previous work \citep[e.g.,][]{Sullivan10, Betoule14, Scolnic18, Kelsey21, Boone21}, although lower and less significant.

\subsection{Further Exploration}
\label{subsec:further_exploration}

The results of the light-curve decomposition in Section~\ref{subsec:lc_decomp} depend on the phase range considered. We explored different ranges with lower limits of $-$8, $-$10 and $-$12 days, and upper limits of $+$12, $+$15 and $+$18 days. As not all SNe have the same coverage, we used an initial sample of 264 SNe in common for all the ranges. Data outside of these phase ranges are usually more incomplete. 50 SNe were discarded for having unphysical-looking light curves, as previously described, leaving 214 SNe. From these 214 SNe, only 156 SNe had \colour values. For details about the discarded SNe at different stages of the analysis process, see Table~\ref{tab:analysis_cuts}.

Our full analysis (Sections~\ref{subsec:lc_decomp} and \ref{subsec:mod_tripp}) was performed for these phase ranges and the Hubble residual r.m.s. values are in Table~\ref{tab:hr_rms}. The range of [$-$10, $+$15]\,d produced the smallest r.m.s. Although naively it might be expected that larger ranges would contain more light-curve information in the NMF components, the fits can also be less reliable at early or late epochs. For example, the ranges starting at $-$12 days produce the largest r.m.s. values, due to less reliable fits from the larger observational uncertainties at these early epochs. A similar behaviour is seen for $+$18\,d. The range of [$-$10, $+$15]\,d produces the best combination between information incorporated and reliable fits.

The shape and information contained in the different NMF components also depends on the number of components chosen. We repeated the analysis using two, four and five components for the phase range of [$-$10, $+$15]\,d. Using two components resulted in a clearly worse result (Hubble residual r.m.s. of $0.141$\,mag), while using four and five components produced similar results as that with three components (both with a Hubble residual r.m.s. of $0.121$\,mag). We conclude that two components are not sufficient to capture the variability of SNe Ia. On the other hand, as the number of components increases it is hard to determine if there is any physical interpretation and/or contribution to the light-curve standardisation. The results suggest that three components is optimal with the current data, each component with clear physical interpretation. Although we did not produce an improved performance over SALT2, despite using one extra parameter, we stress that this is an exploratory work where a single-band light-curve decomposition was performed. Future work with multiple bands may produce improved standardisation as stretch and colour information would be included.

\begin{center}
\begin{table}
\caption{Hubble residual (HR) r.m.s. for our method using different combinations of light-curve phase ranges. The range that gave the lowest r.m.s. is in bold.}
\centering
\begin{tabular}{c|c|c}
Minimum phase & Maximum phase & HR r.m.s.\\
\,[days] & [days] & [mag]\\
\hline
$-$8 & $+$12 & 0.125    \\
$-$8 & $+$15 & 0.120    \\
$-$8 & $+$18 & 0.123    \\
$-$10 & $+$12 & 0.123    \\
\textbf{$-$10} & \textbf{$+$15} & \textbf{0.118}    \\
$-$10 & $+$18 & 0.122    \\
$-$12 & $+$12 & 0.125    \\
$-$12 & $+$15 & 0.125    \\
$-$12 & $+$18 & 0.127    \\
\end{tabular}
\item \textbf{Notes.} The HR r.m.s. for the same sample, using SALT2, is 0.111\,mag.
\label{tab:hr_rms}
\end{table}
\end{center}

To summarise the sample used for the cosmological analysis, of the 620 SNe that \pisco successfully fit (obtained \mb), only 264 have phase coverage in the $B$-band between $-$12 and $+$18 days. There are more SNe with smaller phase-range coverage, but to ensure a fair comparison, we use the sample from the largest phase-range coverage we tested, which has SNe in common with all the other phase ranges. Of these 264 SNe, 50 had unphysical-looking light curves from visual inspection and were removed, leaving 214 SNe on which the NMF analysis was formed. Finally, from these 214 SNe, only 156 have \colour and form our final cosmological sample. We note that, although the cuts in the sample are severe, this is a proof-of-concept analysis with the idea of exploring new approaches and thus requires a \lq golden\rq\ sample, removing possible biases.


\section{\pisco Colour Law}
\label{sec:colour_law}

The SN colour is the light-curve parameter that contributes most to the standardisation of SNe Ia. It is related to the physics, progenitors and environments of SNe Ia, and is a complicated parameter as there are several factors that may contribute to variations in the observed colour (e.g., SN circumstellar material or host galaxy extinction).

The wavelength-dependent variation of colour in SNe Ia is known as the \lq colour-variation law\rq\ (CL). SALT2 describes the CL as a wavelength-dependent function that does not vary in time or as a function of $x_1$ (see equation~(\ref{eq:salt2})). We can also use the \pisco mangling fits to estimate a CL, in a similar fashion to SALT2. The mangling function at \tmax (see Section~\ref{subsec:sed_model}) describes how the colour of a SN varies with respect to a base SED and therefore gives us a method to estimate the relative CL for a single SN (i.e., the mangling function is equal to \colour multiplied by the CL).

We divide SNe into bins of $0.05$\,mag in colour, and calculate their average mangling function in each bin. We then fit a third-order polynomial, optimising across all bins of \colour simultaneously, to obtain a functional form for the CL following the assumption that the CL is wavelength-dependent only (as in equation~(\ref{eq:salt2})). Only data between $\sim$3500 and $\sim$7000\,\angstrom is used as not all SNe have coverage bluer than 3500\,\angstrom, especially at low redshift. High-$z$ SNe get their ultraviolet (UV) wavelengths redshifted to optical wavelengths in the observer frame. However, most of the SNe with rest-frame UV coverage do not have \pisco-measured \colour as the rest-frame $V$ band gets redshifted outside the filters' range, and so are not used for the estimation of the \pisco CL.

\begin{figure}
	\includegraphics[width=\columnwidth]{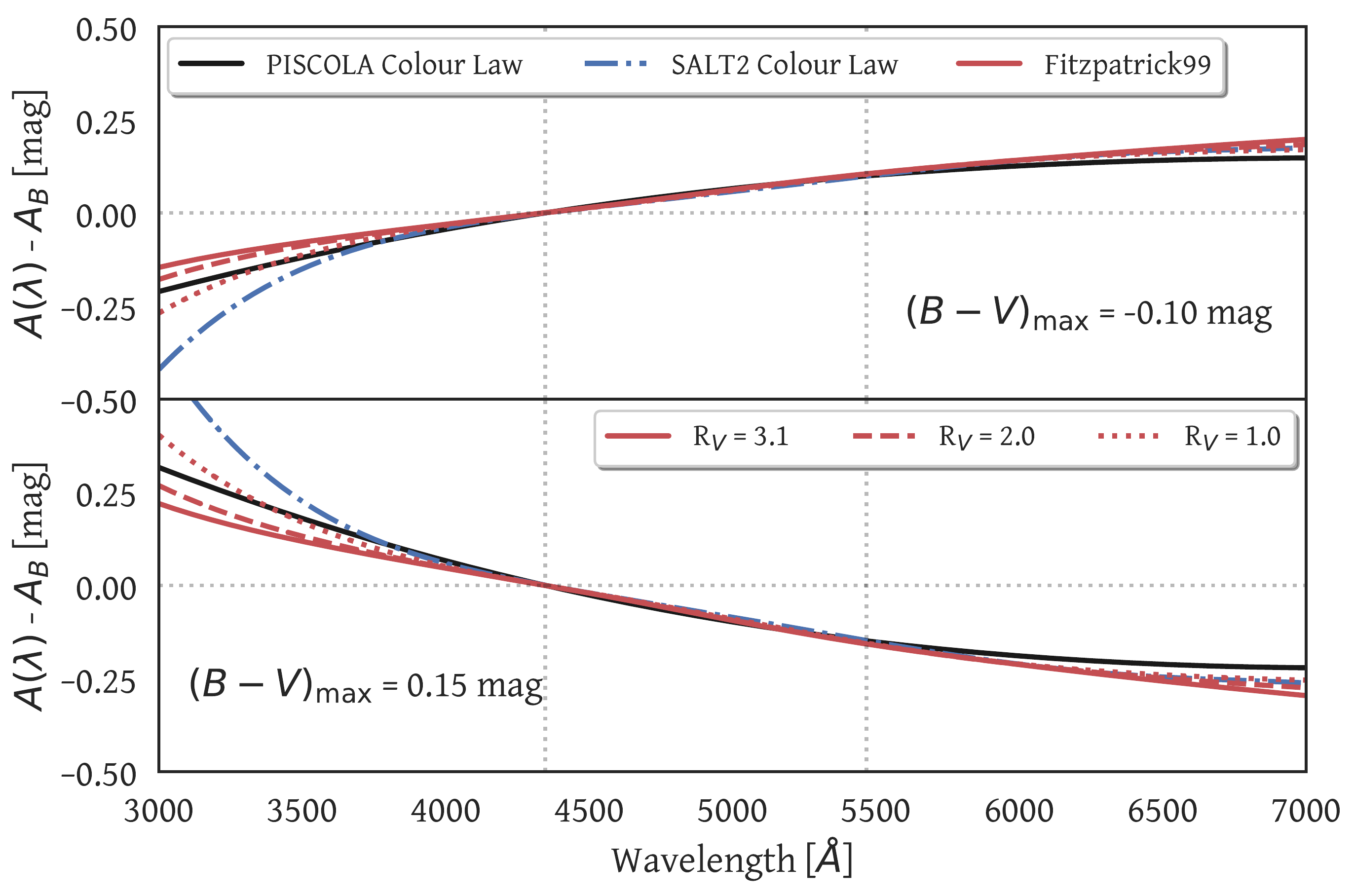}
    \caption{The colour law obtained with \pisco (solid black line) compared to that from SALT2 (dash-dotted blue line) and a \citet[][Fitzpatrick99]{Fitzpatrick99} extinction law for different $R_V$ values (red lines). The vertical dotted lines mark the effective wavelengths of the $B$ and $V$ bands. This comparison is shown for two different \colour values ($-0.10$ and $0.15$\,mag).}
    \label{fig:colour_law}
\end{figure}

In Fig.~\ref{fig:colour_law}, we compare the \pisco CL against that of SALT2 and the extinction law from \citet[][]{Fitzpatrick99} for three different values of the total to selective extinction ratio, $R_V$. The \pisco CL agrees with the \citet[][]{Fitzpatrick99} extinction law with $R_V <\lesssim 3.1$. Similar findings have been reported in previous work \citep[e.g.,][]{Burns14, Amanullah15, Sasdelli16, BS21, Thorp21}.

When comparing with the SALT2 CL, there is agreement at optical wavelengths, but some deviation towards the UV. However, as previously mentioned, the data around 3000\,\angstrom is limited and thus the \pisco CL at these wavelengths is more uncertain.

\begin{figure}
	\includegraphics[width=\columnwidth]{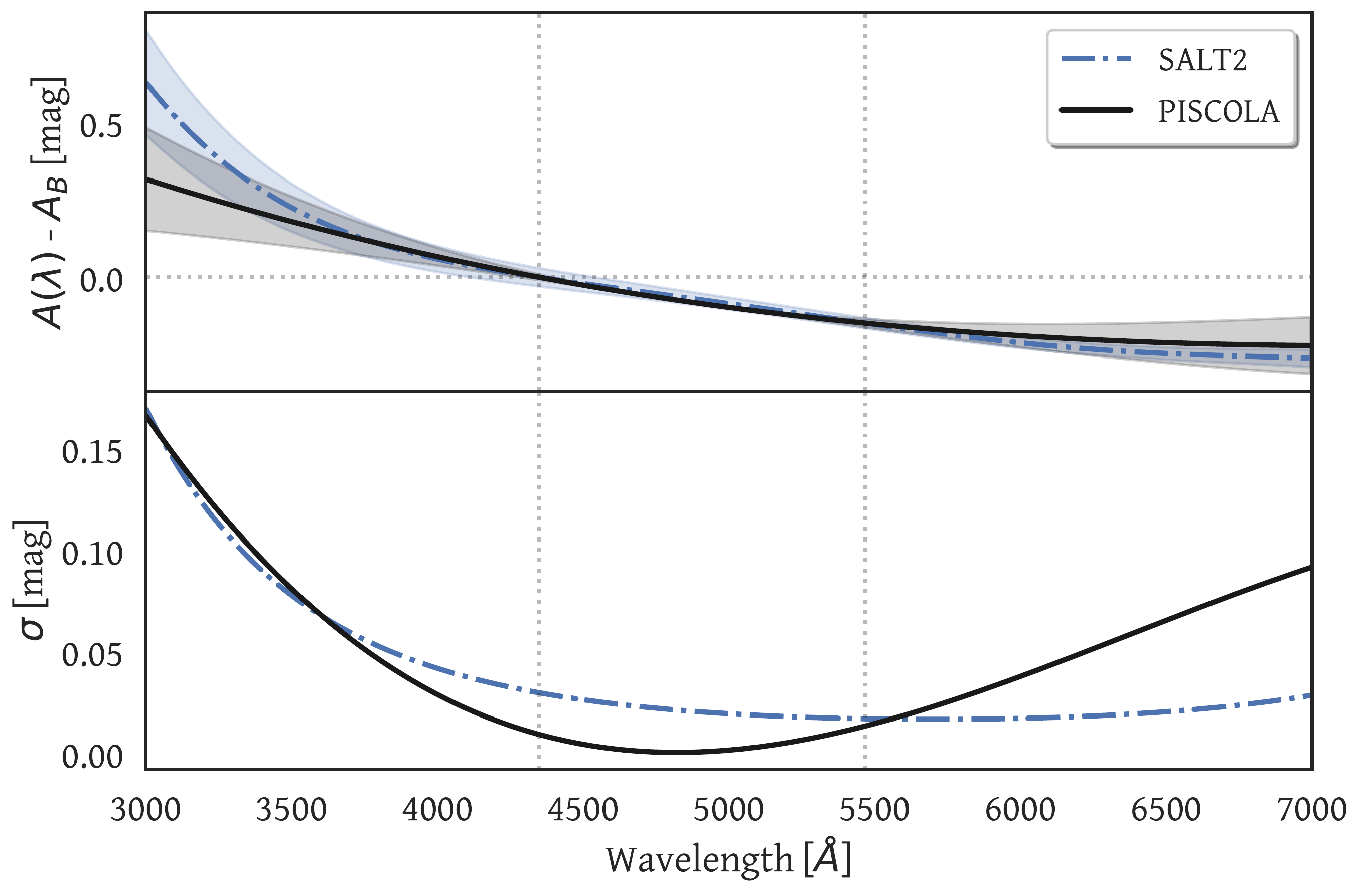}
    \caption{\textbf{Top} panel: CL obtained with \pisco (solid black line) compared to the CL from SALT2 (dash-dotted blue line) for \colour $= 0.15$ mag. The shaded areas represent their respective colour dispersions. \textbf{Bottom} panel: Colour dispersion comparison between \pisco (solid black line) and SALT2 (dash-dotted blue line). The vertical dotted lines mark the effective wavelengths of the $B$ and $V$ bands.}
    \label{fig:colour_disp}
\end{figure}

The colour dispersion, i.e., the scatter around the CL, is also of importance to quantify the disagreement. SALT2 includes its CL in its model (equation~(\ref{eq:salt2})), possibly limiting its behaviour, and it is estimated during the training phase, while we estimate it after the \pisco fitting and correction process of the SNe Ia, almost directly from the data. In Fig.~\ref{fig:colour_disp}, we compare the CLs of \pisco and SALT2 at \colour $= 0.15$\,mag, but also include their respective colour dispersions. We note that, despite the differences in the estimation of the CLs, they agree within the uncertainties, especially in the range used for calculating the \pisco CL (at $\sim$3500--7000\,\angstrom).

\begin{figure}
	\includegraphics[width=\columnwidth]{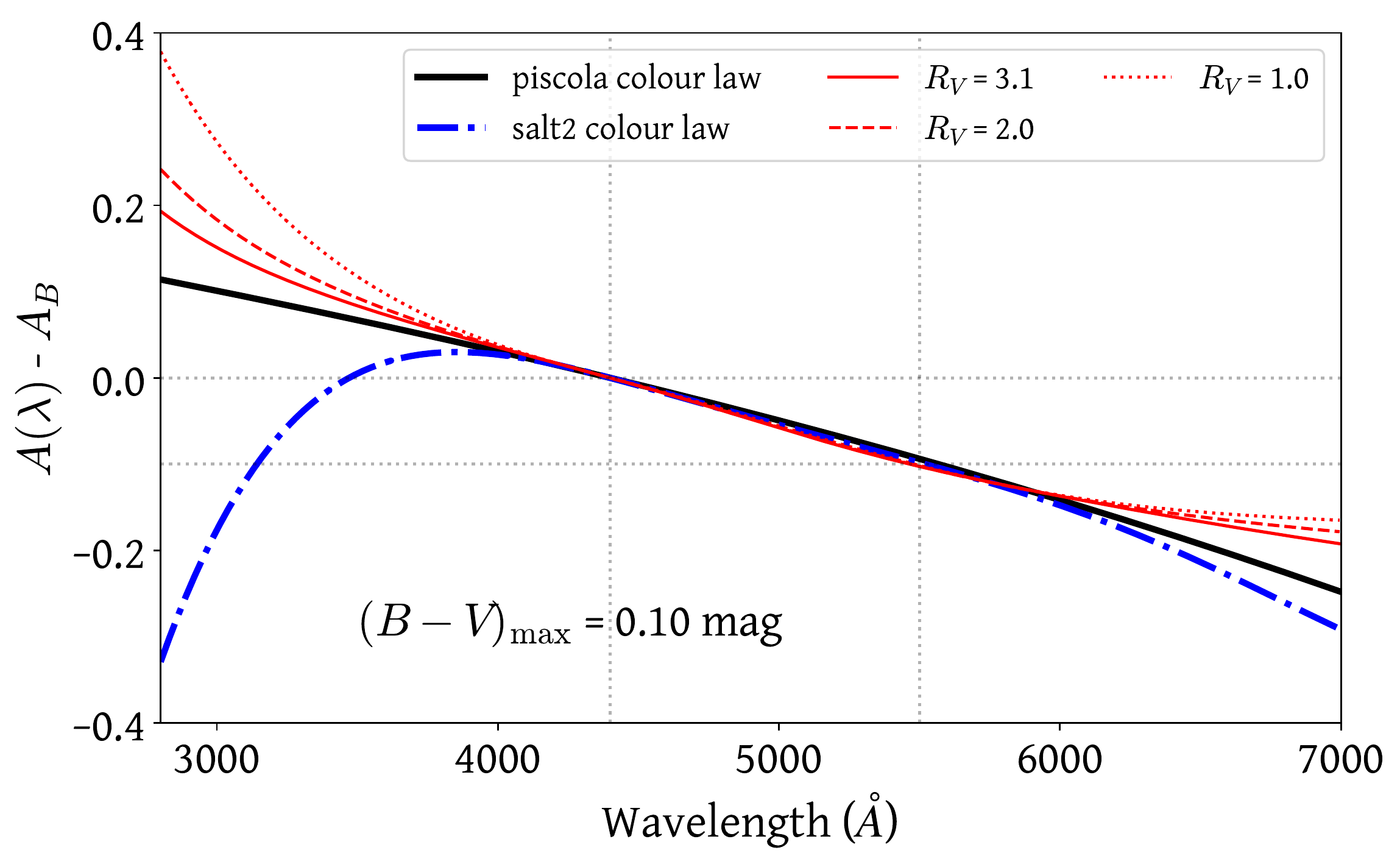}
    \caption{The description is the same as in Fig.~\ref{fig:colour_law} but for \colour$=0.10$\,mag and a SALT2 model with a different (artificial) CL shape. \pisco is able to retrieve a CL that agrees with that of SALT2 (at $\gtrsim$3500\,\angstrom) even after changing the shape of the CL in the SALT2 model.}
    \label{fig:alt_colour_law}
\end{figure}

This was also tested with the simulations of Sec.~\ref{subsec:pantheon_sim}, where we found similar results. Furthermore, by changing the shape of the CL of the simulations, \pisco is able to retrieve a CL that agrees with it (at $\sim$3500--7000\,\angstrom; see Fig.~\ref{fig:alt_colour_law}).

The colour dispersion indicates that \pisco has a larger scatter towards the near-IR, but similar in the UV, compared to SALT2. The shape of the colour dispersion curve from \pisco could be a consequence of a combination of three factors. The first is due to the smaller observational uncertainties in the $B$ and $V$ bands, where much of the flux of a SN Ia emerges, and larger uncertainties at redder and bluer wavelengths. The second is due to the reduced number of SNe covering bluer and redder wavelengths (around 3500 and 7000\,\angstrom respectively) compared to the range around $B$ and $V$. The third is that there may be a difference in the CL for different SNe.

Differences in CLs are important as they can represent differences in the underlying physics of SNe Ia. The SALT2 model mainly incorporates near-UV observations of high-$z$ SNe (e.g., from SNLS) as these wavelengths are redshifted to optical wavelengths and can be observed with ground-based telescopes. In principle, the characteristics of these SNe could be intrinsically different to those at lower redshift \citep[e.g.,][]{Ellis08, Maguire12}. This may introduce biases in the SALT2-measured colour. The CL obtained with \pisco agrees with the average extinction law measured in the MW (with $R_V \lesssim 3.1$) within the uncertainties, implying that the remaining variation (intrinsic scatter) in SNe Ia may be driven by dust of similar properties to that in our galaxy. We note that \citet[][]{BS21} found the opposite, i.e., dust with different properties to that in our galaxy, although using a SALT2 CL.

\section{Summary}
\label{sec:summary}

In this work we have presented a new open-source SN light-curve fitting code \pisco, which relies on Gaussian processes (GP), a data-driven interpolation method, for fitting light-curves. \pisco can be applied to any observer-frame SN light-curves, and produces rest-frame light-curves as its principal output. \pisco can additionally estimate rest-frame light-curve parameters, such as peak magnitudes, colours, and light-curve shapes.

We tested \pisco by applying it to SN Ia data, both simulated and real. With simulations of SNe Ia for different cadences and observational uncertainties, we found that \pisco is reliable for observational cadences of $\lesssim 7$ days for typical current SN Ia samples, provided relatively loose constraints on data coverage around peak luminosity and signal-to-noise are used. When comparing \pisco outputs on real data and comparing to light-curve fits with the SALT2 light-curve fitter, we see small but significant ($>$3$\sigma$) differences in peak rest-frame $B$-band magnitude. However, with no ground-truth for these tests, we argue that such differences may be expected given the different assumptions used in the two techniques.

We then demonstrated a scientific use of \pisco by analysing the rest-frame $B$-band light curves of the Pantheon SN Ia sample using NMF, a machine-learning decomposition algorithm, to search for alternative standardisations of these objects. NMF allows the extraction of \lq easy-to-interpret\rq\ and non-orthogonal components, unlike other algorithms, such as PCA. We compared the NMF coefficients with different SN Ia parameters, and used them to build a Hubble diagram. We tested different combinations of light-curve phase ranges and numbers of components for the decomposition, and found the best results were based on $B$-band light curves with a phase range of [$-10$, $+15$]\,d and three components. This parametrization produces an r.m.s. in the Hubble residual similar to that of SALT2 ($0.118$ and $0.111$ \,mag, respectively), showing the promise of this new framework. We stress that, although \pisco does not outperform SALT2 in this particular case, future work with multiple bands can produce better results. Additionally, this analysis is only a limited demonstration of \pisco 's potential.

\pisco uses a smooth GP interpolation to adjust its base SED to an observed SN colour (a mangling function). This mangling function encodes information on the colour law (CL) of SNe Ia. We estimated a functional form for this CL by fitting a third-order polynomial and compared it with the SALT2 CL and \citet{Fitzpatrick99} extinction laws with different $R_V$ values. We found that the \pisco CL agrees with an extinction law with $R_V \lesssim 3.1$, but also with the SALT2 CL. Although there could be some slight disagreement towards the UV, a possible cause is the extrapolation of the \pisco CL bluer than $\sim$3500\,\angstrom.

We have plans for future upgrades of \pisco. This includes the use of a \lq stretch\rq-dependent time-series SED, a mangling function in two-dimensions (wavelength and time), and exploring alternative GP models, for example the use of different kernels, different bounds for the hyper-parameters, etc.. This may produce more accurate light-curve fits and mangling functions. Finally, we emphasise that although our tests have been based around applications to SNe Ia, \pisco is generic so that, with an appropriate time-series SED for K-corrections, it can fit any type of optical transient to estimate rest-frame light-curves and luminosities. For instance, SED templates for type Ib, Ic and II SNe already exist \citep[e.g.,][]{Vincenzi19} and, in the case of superluminous SNe, a black-body could be used as an approximate SED template as no current templates exist.

\section*{Acknowledgements}

\textit{Software}: \\
\texttt{numpy} \citep{harris2020array}, 
\texttt{matplotlib} \citep{matplotlib}, 
\texttt{seaborn} \citep{seaborn},
\texttt{pandas} \citep{pandas},
\texttt{scipy} \citep{scipy},
\texttt{lmfit} \citep{lmfit},
\texttt{scikit-learn} \citep{scikit-learn}, 
\texttt{emcee} \citep{emcee}, 
\texttt{coner} \citep{corner}, 
\texttt{george} \citep{Ambikasaran16}, 
\texttt{sfdmap}\footnote{\url{https://github.com/kbarbary/sfdmap}}, 
\texttt{extinction}\footnote{\url{https://github.com/kbarbary/extinction}} \citep{Barbary2016},
\texttt{peakutils}\footnote{\url{https://github.com/lucashn/peakutils/tree/v1.1.0}} \citep{peakutils},
\texttt{sncosmo} \citep{sncosmo}.\\

TMB acknowledges funding from the CONICYT PFCHA/DOCTORADOBECAS CHILE/2017-72180113. We thank Dalya Baron for discussions regarding the use and understanding of principal component analysis and similar algorithms. TMB would also like to thank the LSSTC-DSFP which allowed him to develop computational skills and tools for the development of \pisco and the analysis in this work. We thank Santiago Gonz\'alez-Gait\'an and Llu\'is Galbany for discussions about light-curve fitting and decomposition. Finally, we thank Dan Scolnic for his patience in helping us understand the intricacies of the calibration of the Pantheon SN Ia sample.

\section*{Data Availability}

All the data used in this work was publicly available. \pisco is an open source code and can be found in a Github repository: \url{https://github.com/temuller/piscola}.


\bibliography{References}
\bibliographystyle{mnras}


\appendix


\section{Gaussian process}
\label{app:gp}

A GP is a random process where any point $\mathbf{x} \in \mathbb{R}^{d}$ is assigned a random variable $f(\mathbf{x})$ and where the joint distribution of a finite number of these variables $p\left(f\left(\mathbf{x}_{1}\right), \ldots, f\left(\mathbf{x}_{N}\right)\right)$ is itself Gaussian: $ p(\mathbf{f} \mid \mathbf{X})=\mathcal{N}(\mathbf{f} \mid \boldsymbol{\mu}, \mathbf{K}) $, where $\boldsymbol{\mu}$ and $\mathbf{K}$ are the mean and covariance functions. The latter is also called \textit{kernel}.

\begin{figure}
	\includegraphics[width=\columnwidth]{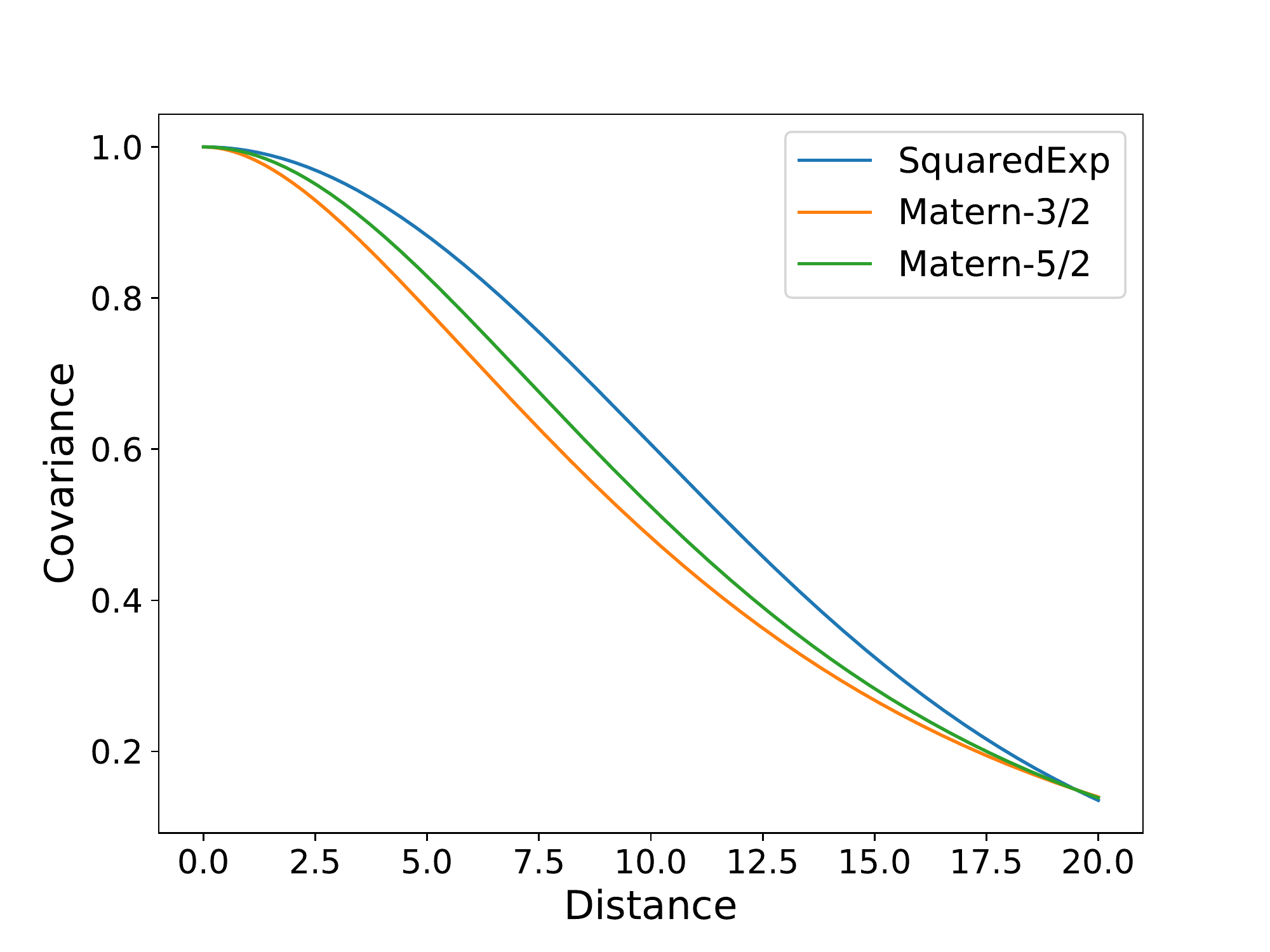}
    \caption{Comparison of the covariance between three kernels available with \pisco: \se (blue), \mtt (orange) and \mft (green).}
    \label{fig:covariance_kernels}
\end{figure}

\pisco incorporates three different kernels: Squared Exponential, \mtt\ and \mft. These are defined as
\begin{equation}
    k_{SE}(x, x')  = \sigma^2 \text{exp} \Bigg( -\frac{\lvert x - x' \rvert^2}{2\ell} \Bigg),
\end{equation}
\begin{equation}
    k_{M32}(x, x') = \sigma^2 \Bigg( 1 + \frac{\sqrt{3}\,\lvert x - x' \rvert}{\ell} \Bigg) \text{exp} \Bigg( -\frac{\sqrt{3}\,\lvert x - x' \rvert}{\ell} \Bigg),
\end{equation}    
and
\begin{equation}
\begin{split}
    k_{M52}(x, x') = \sigma^2 \Bigg( 1 + \frac{\sqrt{5}\,\lvert x - x' \rvert}{\ell} + &\frac{5\,\lvert x - x' \rvert}{3 \ell^2} \Bigg)\\
    &\text{exp} \Bigg( -\frac{\sqrt{5}\,\lvert x - x' \rvert}{\ell} \Bigg),
\end{split}
\end{equation}
where $\ell$ (length-scale) and $\sigma^2$ (variance) are the hyperparameters of the kernels. Fig.~\ref{fig:covariance_kernels} shows a comparison between the three kernels. From the mathematical perspective, each kernel has its own properties; however, in terms of applications for fits, the main difference is the covariance between points, which is larger for the \se kernel than for the other two, assuming two data points at a fixed distance, and a fixed set of hyperparameters. The \textit{Mat\'ern} family is appropriate for modelling physical processes (the \se kernel can be too smooth), with the \mtt and \mft kernels of special interest as they are simple functions.

\begin{figure}
	\includegraphics[width=\columnwidth]{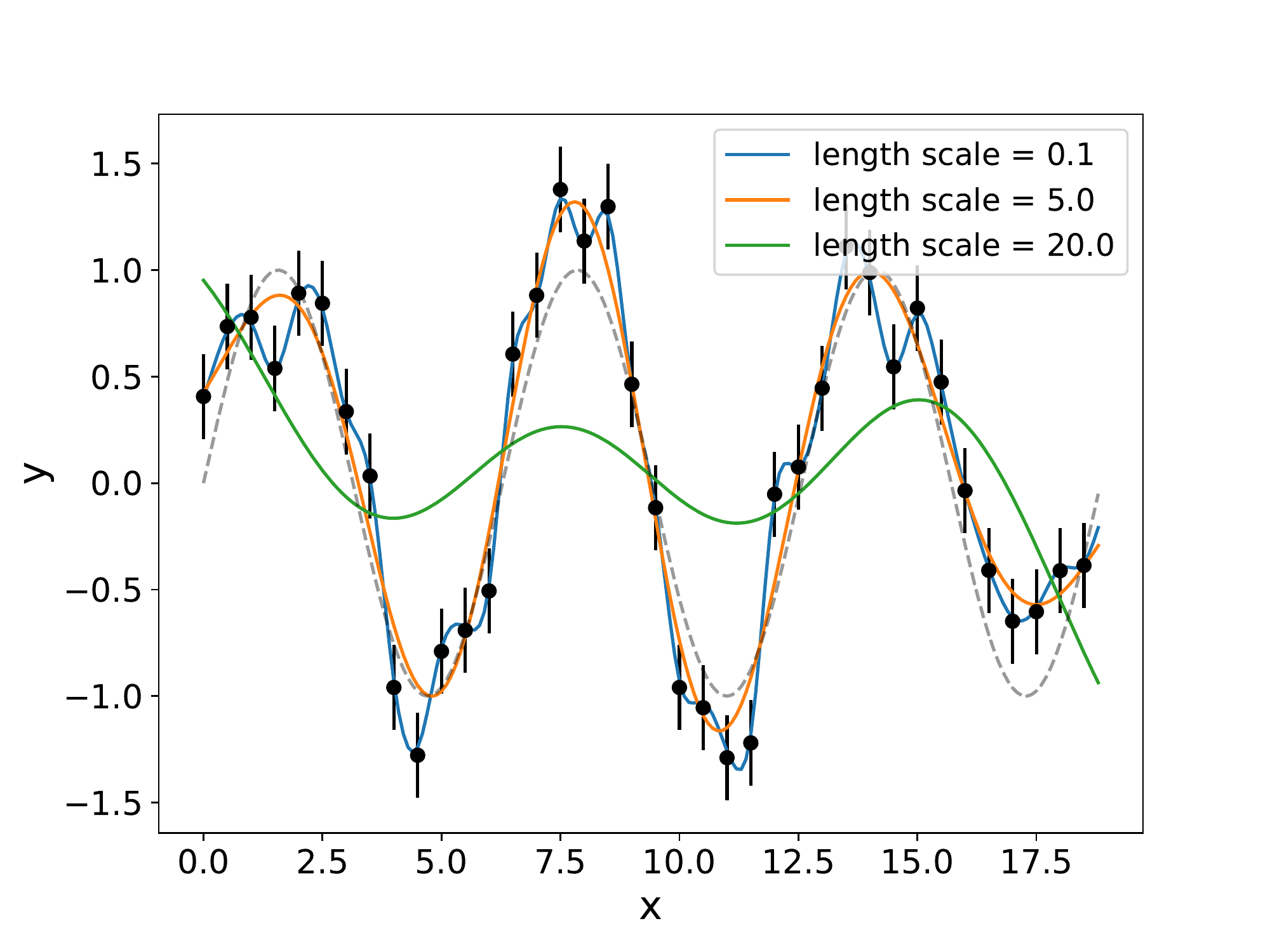}
    \caption{Comparison between different length scale values ($\ell$) for an \se kernel: 0.1 (blue), 5.0 (orange) and 20.0 (green). GP fits (solid lines) were performed on simulated data (black circles). Only the mean GP model is shown in each case for visualisation. The underlying function, from which the simulated data was extracted from, is also  shown (dashed grey line).}
    \label{fig:kernels_length_scale}
\end{figure}

\begin{figure}
	\includegraphics[width=\columnwidth]{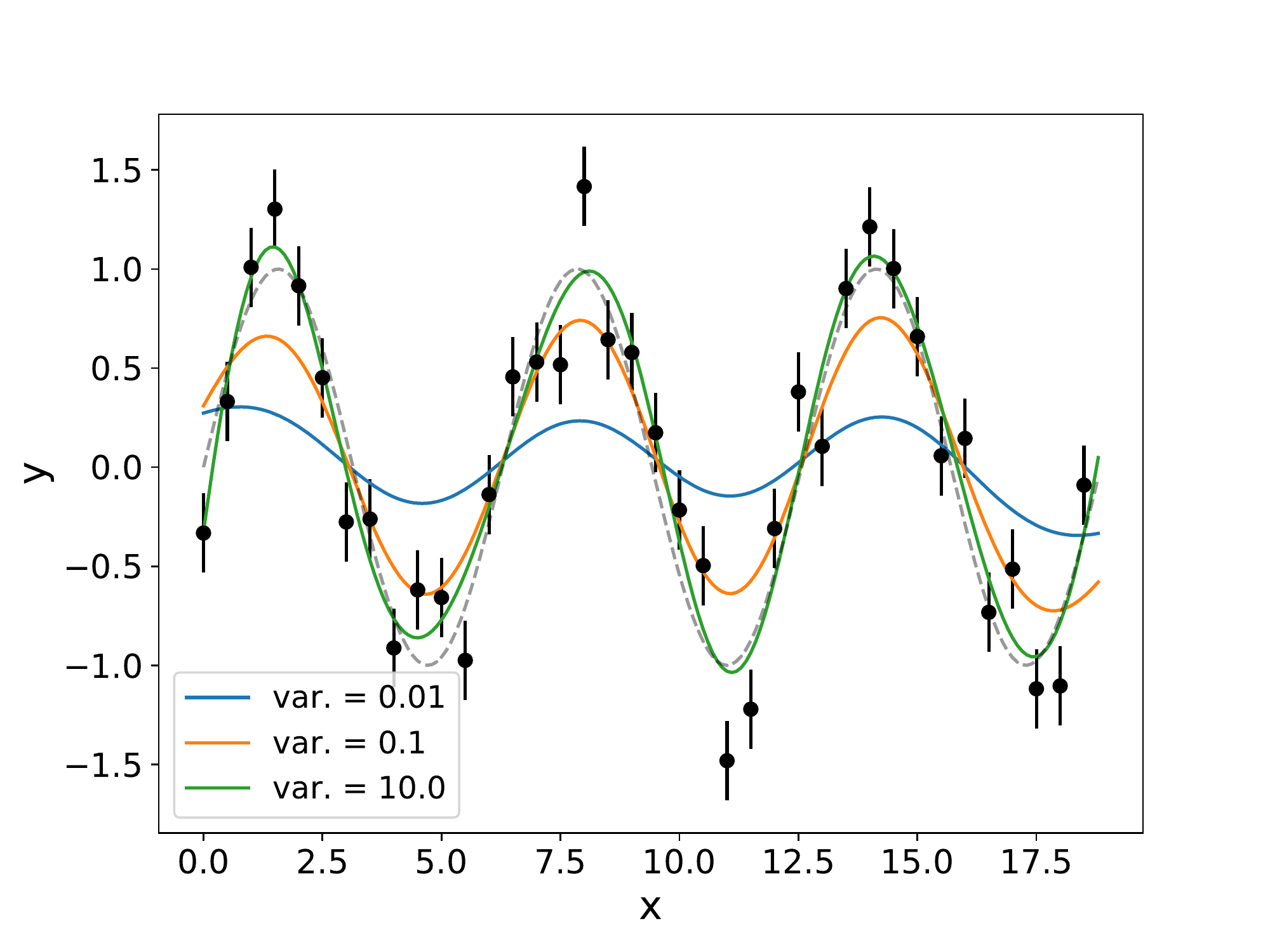}
    \caption{Comparison between different variance values ($\sigma^2$) for an \se kernel: 0.01 (blue), 0.1 (orange) and 10.0 (green). GP fits (solid lines) were performed on simulated data (black circles). Only the mean GP model is shown in each case for visualisation purposes. The underlying function, from which the simulated data was extracted from, is also shown (dashed grey line).}
    \label{fig:kernels_variance}
\end{figure}

Figs.~\ref{fig:kernels_length_scale} and \ref{fig:kernels_variance} show a comparison between different length scales and variances for the \se kernel. The length scale determines the length of the \lq wiggles\rq\ in a function, giving a sense of how far the data can be extrapolated. The variance determines the average distance of the function away from its mean. For a detailed introduction to GP method see \cite{Rasmussen06}. 

\subsection{One dimensional Gaussian process fits}
\label{app:1d_gp}

\begin{figure}
	\includegraphics[width=\columnwidth]{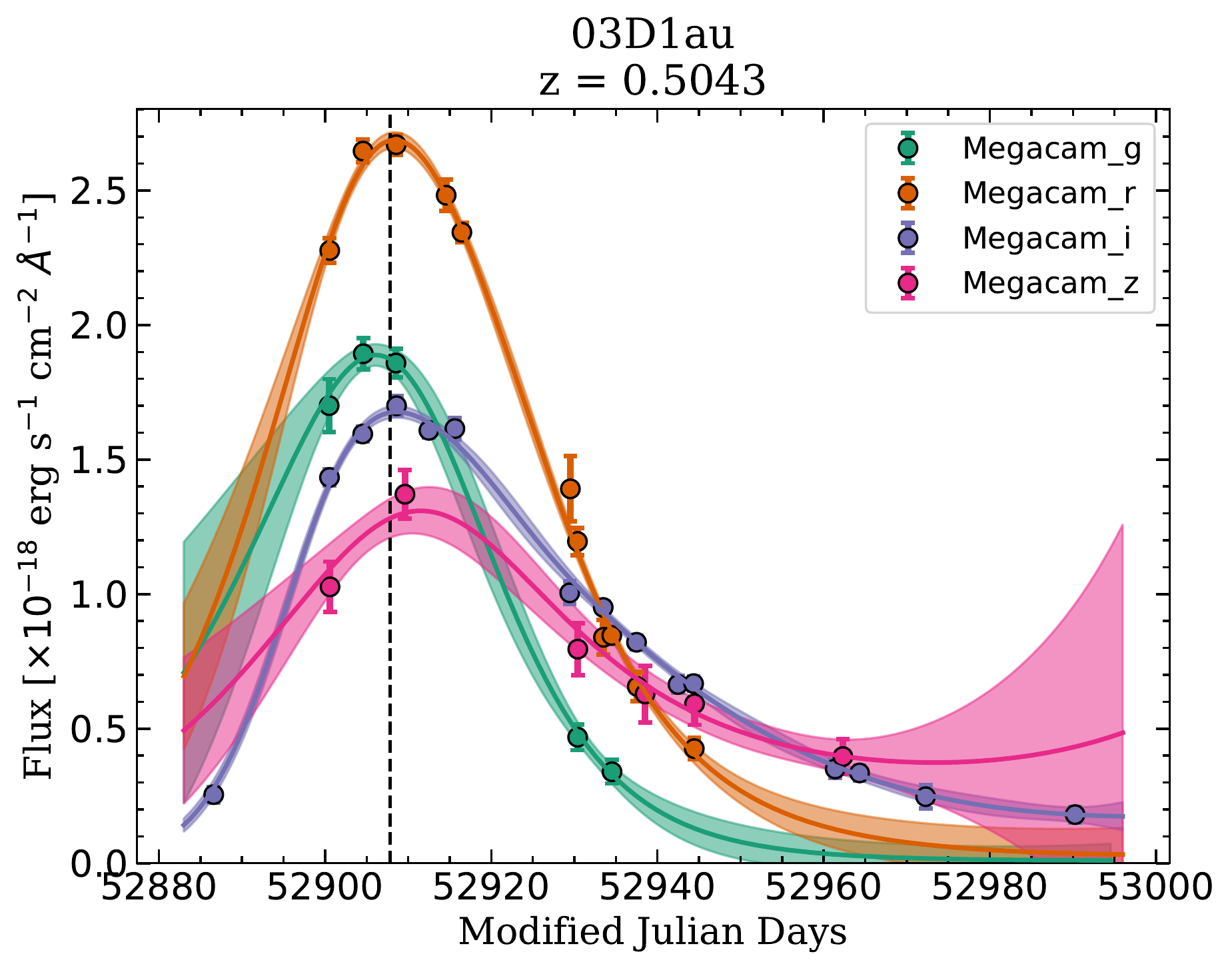}
    \caption{The description is the same as in Fig.~\ref{fig:lc_fits}, but for a 1D GP model, with independent fits for each band.}
    \label{fig:lc_fits_1D}
\end{figure}

Two-dimensional GP light-curve fits have the advantage of performing informative interpolation/extrapolation given data from multiple bands and this forms our default approach. However, we also explored fitting each filter independently (a 1D GP model). In Fig.~\ref{fig:lc_fits_1D}, we show the same example as in Fig.~\ref{fig:lc_fits}, but for such a 1D GP model, with independent fits for each band, i.e. a regression of flux as a function of time. The extrapolation of the rise of the light curve for the $r$ and $z$ bands over-predicts the flux by comparing them with $i$ band, given that the former bands are bluer and redder than the latter. In addition, the uncertainty of the $z$ light-curve decline fit rapidly increases the further the GP model extrapolates. A final consideration is the estimation of \tmax: as in a 1D fit there is no interpolation in the wavelength axis, the peak can only be estimated by choosing the closest observer-frame band to rest-frame $B$ band or using a redder and bluer band than the $B$ band for a more accurate estimation. Both approaches are less precise than the estimation from a two-dimensional interpolation.

\subsection{Effect of S/N}
\label{app:gp_cad_unc}

\begin{figure}
	\includegraphics[width=\columnwidth]{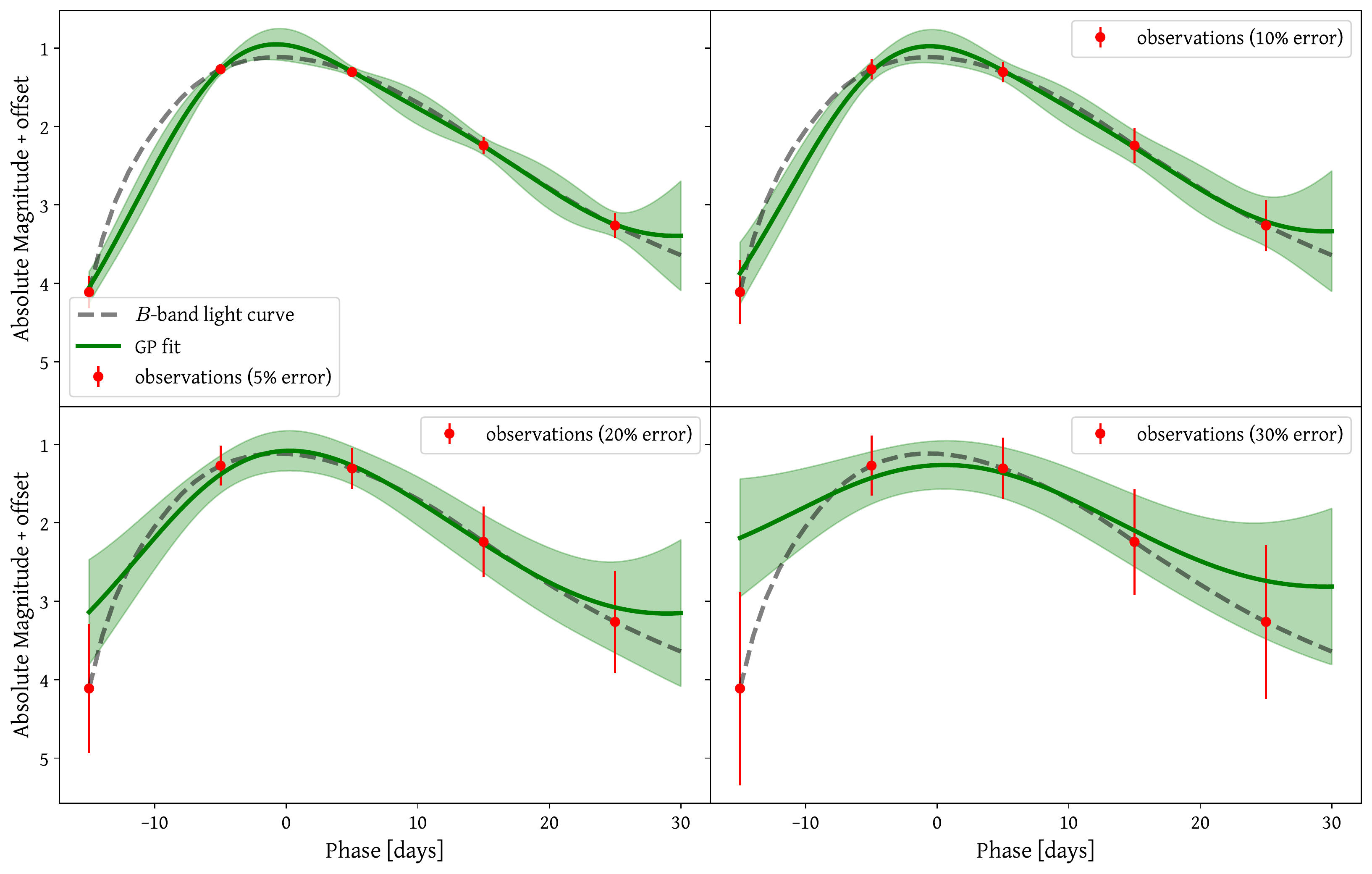}
    \caption{10-day cadence 'observations' (red circles) extracted from the rest-frame $B$-band light curve (dashed grey line) of a SN Ia SED template. GP fits (in green) were performed for cases with simulated fractional errors of the observations of 5, 10, 20 and 30 per cent.}
    \label{fig:cad_and_unc_effect}
\end{figure}

The results of a GP fit depends on the cadence and uncertainties of the observations. In Fig.~\ref{fig:cad_and_unc_effect} we show a GP fit to a rest-frame $B$-band light curve taken from a SN Ia SED from which we extracted \lq observations\rq\ with a 10-day cadence. We simulate fractional errors of the observations with 5, 10, 20 and 30 per cent values, reproducing the effect of S/N. When the errors are small (5 and 10 per cent), the peak is over-predicted (e.g., for low-$z$ SNe). Conversely, when they are large (30 per cent), the peak is under-predicted (e.g., for SDSS SNe).

This just gives a general idea on how S/N can affect the light-curve fits as the exact values of the uncertainties will depend on the apparent magnitude of the SN. For example, for two SNe with peak apparent magnitudes of 17 and 14, a 1 per cent error turns into 0.17\,mag and 0.14\,mag, respectively.

\section{Light-Curve Fits}
\label{app:lc_fits}

\begin{figure}
	\includegraphics[width=\columnwidth]{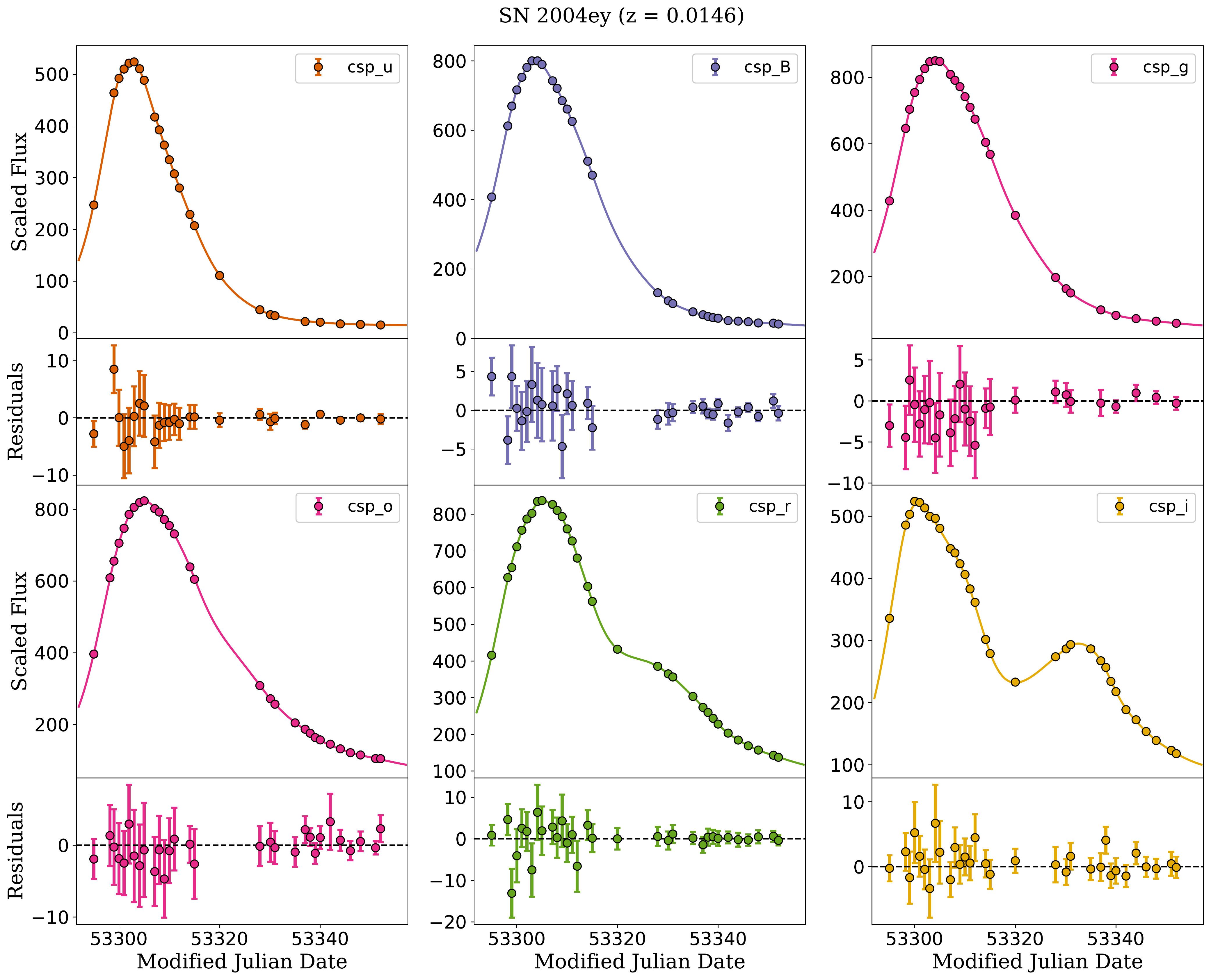}
    \caption{\pisco light-curves fit and residuals of SN 2004ey from the CSP survey. The observational uncertainties for this SN are small so they are only seen in the residual plots. The label \texttt{csp\_o} is used by \pisco to refer to one of the three available CSP $V$-bands.}
    \label{fig:sn_fit_pisco}
\end{figure}

\begin{figure}
	\includegraphics[width=\columnwidth]{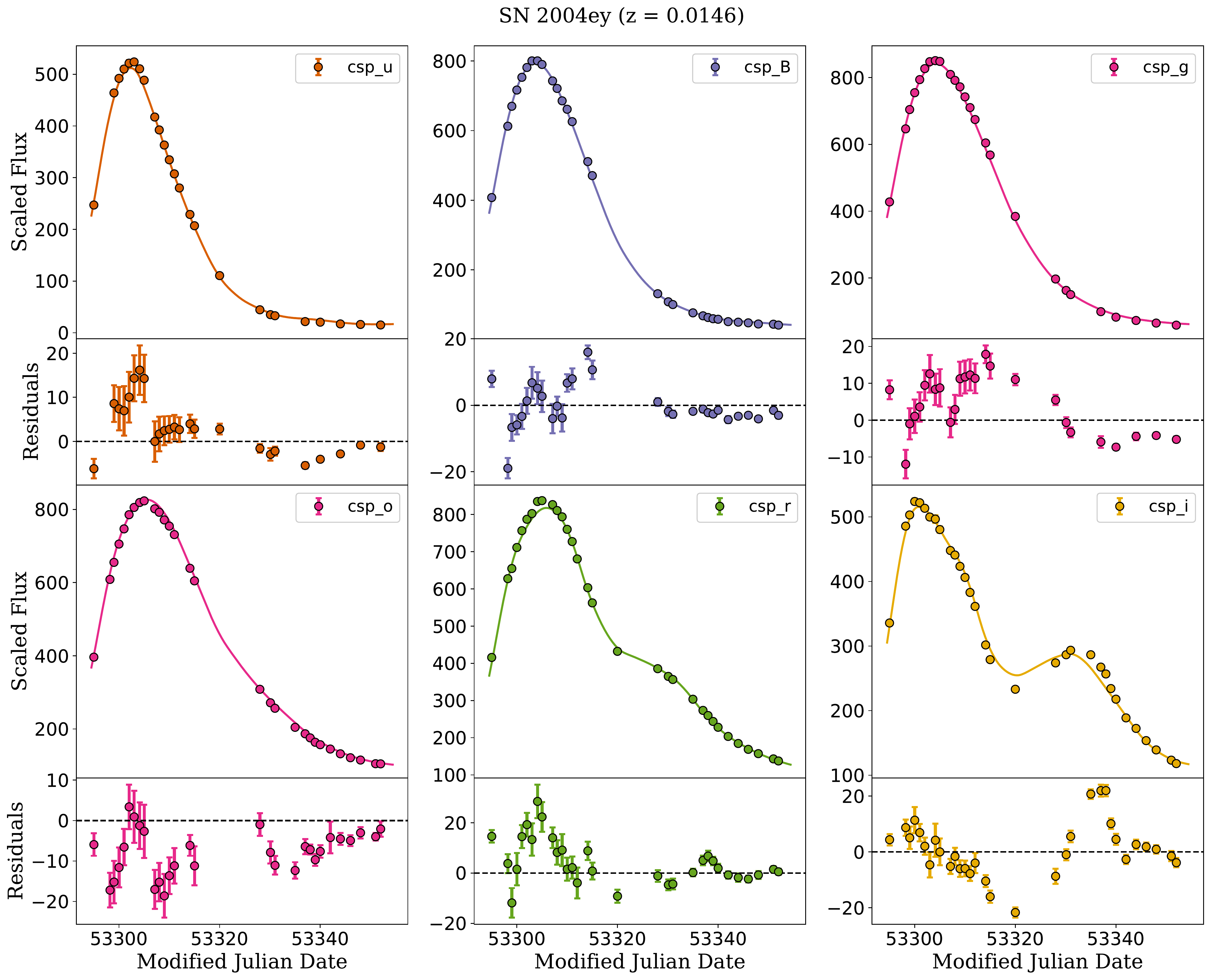}
    \caption{SALT2 light-curves fit and residuals of SN 2004ey from the CSP survey. The rest of the description is the same as in Fig.~\ref{fig:sn_fit_pisco}.}
    \label{fig:sn_fit_salt2}
\end{figure}

\pisco presents several advantages over template-driven light-curve fitters, one of them being the better fits of low-$z$ SNe, i.e., objects with relatively well-sampled light curves and high S/N. One example is presented in Fig.~\ref{fig:sn_fit_pisco}, where we show the \pisco fit to SN 2004ey from the CSP survey. In Fig.~\ref{fig:sn_fit_salt2}, we show the SALT2 fit we performed to the same SN, using the implementation of the code in \texttt{sncosmo}. In this case, we see that SALT2 produces larger residuals compared to \pisco.

\pisco outperforms SALT2 when fitting the CSP $u$ and $r$, which are on the blue and red limits, respectively, of the effective training range of SALT2 ($\sim$3000--7000\,\angstrom). Furthermore, \pisco produces good results in the $i$ band, even fitting the secondary peak, showing the potential of fitting NIR light curves. We also note that \pisco produces better results in bands like CSP $g$ and $V$ (\texttt{csp\_o} in Fig.~\ref{fig:sn_fit_salt2}), important for the estimation of colour and colour evolution in SNe Ia.

Future work will involve using \pisco to fit NIR bands of SNe Ia and estimate distances with them.

\bsp	
\label{lastpage}
\end{document}